\documentclass[fleqn,usenatbib]{mnras}

\usepackage{newtxtext,newtxmath}

\usepackage[T1]{fontenc}
\usepackage{ae,aecompl}

\usepackage{graphicx}	
\usepackage{amsmath}	
\usepackage{amssymb}	
\usepackage{booktabs,multirow}

\newcommand{\sectionname}{Section}
\newcommand{\figname}{Figure}
\newcommand{\eqname}{Equation}

\newcommand{\project}[1]{\textsl{#1}}
\newcommand{\acronym}[1]{{\small{#1}}}
\newcommand{\gaia}{\project{Gaia}}

\newcommand{\hipparcos}{\project{Hipparcos}}

\newcommand{\tgas}{\acronym{TGAS}}
\newcommand{\gdrtwo}{\acronym{{GDR2}}}
\newcommand{\bprp}{\ensuremath{\mathrm{BP}-\mathrm{RP}}}
\newcommand{\feh}{\text{[Fe/H]}}

\newcommand{\kms}{\ensuremath{\rm km~s^{-1}}}
\newcommand{\msun}{\ensuremath{{\rm M}_\odot}}
\newcommand{\pc}{{\rm pc}}
\newcommand{\given}{\,|\,}

\newcommand{\bs}[1]{\boldsymbol{#1}}

\newcommand{\mat}[1]{\mathbf{#1}}
\renewcommand{\vec}[1]{\bs{#1}}
\newcommand{\diag}{\ensuremath{\mathrm{diag}}}

\newcommand{\mspc}{\ensuremath{\mathrm{m}\,\mathrm{s}^{-1}\,\mathrm{pc}^{-1}}}
\newcommand{\masyr}{\ensuremath{{\mathrm{mas}\,\mathrm{yr}^{-1}}}}

\newcommand{\fmem}{\ensuremath{f_{\mathrm{mem}}}}
\newcommand{\parallax}{\ensuremath{\pi}}
\newcommand{\pmra}{\ensuremath{\mu_\alpha}}
\newcommand{\pmdec}{\ensuremath{\mu_\delta}}
\newcommand{\Cova}{\ensuremath{\mat{C}_{\vec a, i}}}
\newcommand{\Mnow}{M_{\rm now}}
\newcommand{\tnow}{t_{\rm now}}
\newcommand{\trnow}{t_{\rm r,now}}
\newcommand{\etanow}{\ensuremath{\eta_{\rm now}}}
\newcommand{\rt}{r_{\rm t}}

\graphicspath{{./}{figures/}}

\title[Death Throes of The Hyades]{
  Kinematic modelling of clusters with Gaia:\\
  the Death Throes of the Hyades
}

\author[Oh \& Evans]{
	Semyeong Oh$^{1}$\thanks{Email: soh,nwe@ast.cam.ac.uk} and N. Wyn Evans$^{1}$
	\\
	$^{1}$Institute of Astronomy, University of Cambridge, Madingley Rd, Cambridge, CB3 0HA, UK\\
}

\pubyear{2020}

\begin{document}
\label{firstpage}
\pagerange{\pageref{firstpage}--\pageref{lastpage}}
\maketitle

\begin{abstract}
  The precision of the \gaia\ data offers a unique opportunity to study the
  internal velocity field of star clusters.
  We develop and validate a forward-modelling method for the internal motions
  of stars in a cluster.
  The model allows an anisotropic velocity dispersion matrix and
  linear velocity gradient describing rotation and shear, combines radial
  velocities available for a subset of stars, and accounts for contamination
  from background sources via a mixture model.
  We apply the method to \gaia\ DR2 data of the Hyades cluster and its
  tidal tails,
  dividing and comparing the kinematics of stars within and beyond $10$~pc,
  which is roughly the tidal radius of the cluster.
  While the velocity dispersion for the cluster is nearly isotropic,
  the velocity ellipsoid for the tails is clearly elongated with
  the major axis pointing towards the Galactic centre.
  We find positive and negative expansion at $\approx 2\sigma$ significance
  in Galactic azimuthal and vertical direction for the cluster but no rotation.
  The tidal tails are stretching in a direction tilted from the Galactic
  centre while equally contracting as the cluster in Galactic vertical direction.
  The tails have a shear ($A$) of $16.90\pm 0.92$ \mspc and a vorticity ($B$)
  of $-6.48\pm1.15$ \mspc, values distinct from the local Oort constants.
  By solving the Jeans equations for flattened models of the Hyades, we show
  that the observed velocity dispersions are a factor of $\approx 2$ greater
  than required for virial equilibrium due to tidal heating and disruption.
  From simple models of the mass loss, we estimate that the Hyades is close
  to final dissolution with only a further $\lesssim 30$~Myr left.
\end{abstract}




\section{Introduction} \label{sec:intro}

Stars are born in supersonically-turbulent, self-gravitating Giant Molecular
Clouds, which are hierarchically structured. Dense regions within a
cloud, often referred to as ``clumps'', are thought to be birth sites of star
clusters \citep[][and references therein]{krumholz-cluster-review}.
Although both the molecular clouds and the clumps within them are
gravitationally bound \citep{heyer-dame-molecular-clouds,urquhart2018-clumps},
only $\lesssim 10$\% of the initial stellar groups observed in the
distribution of young stellar objects
eventually become bound star clusters after gas removal
\citep{ladalada2003}.
The pathway by which stellar groups emerging from the hierarchical structure
of a cloud end up as bound clusters is not well understood.

Once formed, the evolution of bound star clusters is governed by both
internal processes, such as stellar and binary evolution, mass segregation
and two body relaxation effects, as well as external influences, such as
galactic tides, bulge and disc shocking, and gravitational interactions with
passing molecular clouds. The complex interplay between internal and external
effects leads to the death throes of many star clusters. For example, tidal
disruption may be enhanced by stellar evolution, leading to mass loss through
winds and supernova explosions. This reduces the density in a cluster and
makes it more fragile to the buffetings of external tidal forces.

Studies of internal kinematics of clusters and associations can provide
critical and direct insights into their formation and evolution, which are
orthogonal to existing constraints such as morphology and stellar
demographics. As internal motions are differences of velocities with respect
to the mean motion and star clusters typically have velocity dispersions (a
proxy for the magnitude of internal motions) of less than a few \kms, their
study requires high-precision astrometric measurements in order to pin down
the positions, velocities and membership of clusters. Until recently, such
studies were limited to a few nearby star-forming regions and specifically
targeted surveys of clusters.

The \gaia\ mission \citep{gaia-mission} has changed the situation
dramatically, as it delivers astrometry for over a billion sources brighter
than $G=20$ and radial velocities for a subset of brighter stars ($G<12$).
Indeed, the \gaia\ second data release \citep[][\gdrtwo]{gaia-dr2-overview}
has already fueled a number of studies on the internal kinematics of clusters
and associations.

Indeed, many recent studies have already examined the internal kinematics of young
($\lesssim 30$~Myr) associations in large star-forming
complexes~\citep{zari2019-orion,kim2019-onc,kuhn2019,kounkel2018-orion,kos2019-orion,
wright_mamajek2018-ScoCen,cantat-gaudin2019-VelaPuppis,wright2019-Lagoon}
with particular interest in the role of gas expulsion from stellar feedback.
The results are varied from subgroups in the Scorpius-Centaurus OB association
showing no expansion \citep{wright_mamajek2018-ScoCen}
to groups in the Vela Puppis region and the Lagoon Nebula showing anisotropic
expansion \citep{cantat-gaudin2019-VelaPuppis,wright2019-Lagoon}.
An emerging picture is that star formation in these regions produces highly
substructured and complex distribution of young stars, and involves multiple
episodes of star forming events.

Here, we present a forward-modelling approach to the kinematic modelling of
clusters suited for taking full advantage of the combination of the \gaia\
astrometry with radial velocities from various spectroscopic surveys. Our
method builds upon and extends the work of \citet{lindegren2000}, which was
developed to infer the astrometric radial velocities from the \hipparcos\
data \citep[see also][for its applications to the \gaia\
data]{reino2018-tgas-hyades,bravi2018,zari2019-orion}. We describe and
justify each components of our model, and validate our implementation with
mock data generated according to the model in \sectionname~\ref{sec:method}.

We apply the method to the specific example of the Hyades cluster in \sectionname~\ref{sec:application}. This is one of the nearest ($d\approx 46$~pc), large ($N \gtrsim 500$ for $G \lesssim 20$), old \citep[$\approx 680$~Myr;][]{gossage2018} open clusters. Historically, the Hyades cluster played an important role as a calibrator of the absolute magnitude-spectral type and the mass-luminosity relation. Modern interest in the Hyades is focused on the cluster's birth, life and death in the Galactic environment. Its proximity to the Sun and thus the availability of high quality astrometric data has stimulated a number of kinematical studies, 
from \hipparcos\ ~\citep{perryman1998,Br01} to the first \gaia\ data release~\citep{reino2018-tgas-hyades}. Very recently, \gdrtwo\ has revealed the existence of tidal tails of the cluster out to $\lesssim 150$~pc from the cluster centre \citep{meingast2019,roser2019}. \sectionname~\ref{sec:models} builds steady-state and evolving models of the Hyades cluster in the light of our kinematic investigations. We discuss the results in the context of complete tidal dissolution of the cluster.

\section{Method} \label{sec:method}

\begin{table*}
  \centering
  \begin{tabular}{llll}
  \hline
  Parameter                 & Description                                                      & Prior                                          & Unit                               \\ \hline
  $N$                       & total number of stars in the sample                              &                                                &                                    \\
  $N_\mathrm{RV}$           & total number of stars with radial velocities                     &                                                &                                    \\
  $i$                       & star index, $i=1,\dots,N$                                        &                                                &                                    \\ \hline
  $\vec{v}_0$               & mean velocity vector $(v_x,\,v_y,\,v_z)$ in ICRS                 & $\mathcal{N}(0,\,50)$                          & \kms                               \\
  $\mat{\Sigma}$            & \multicolumn{2}{l}{(3, 3) velocity dispersion matrix of the cluster, $\Sigma=\diag(\vec{\sigma}_v)\,\mat{\Omega}\,\diag(\vec{\sigma}_v)$}  & $\mathrm{km}^2\,\mathrm{s}^{-2}$\\
  $\vec{\sigma}_v$          & scale vector of $\Sigma$                                         & $\text{Half-Cauchy}(0,\,2.5)$                  & \kms                               \\
  $\mat{\Omega}$            & correlation matrix of $\Sigma$                                   & $\mathrm{LKJ}(2)$                              &                                    \\
  $\mat{T}$                 & velocity gradient matrix, $T_{ij} = d v_i / d x_j$                 & $\mathcal{N}(0,\,50)$                          & \mspc                              \\
  \fmem                     & fraction of the sample that are cluster members                  & $\mathcal{U}(0,\,1)$                           &                                    \\
  $\sigma_\mathrm{bg}$      & velocity dispersion of the background                            & $\mathcal{N}(30,\,20)$, $\sigma_\mathrm{bg}>0$ & \kms                               \\
  $\vec{v}_{0,\mathrm{bg}}$ & mean velocity of the background                                  & $\mathcal{N}(0,\,50)$,                         & \kms                               \\
  $d_i$                     & distance of star $i$                                             &                                                & pc                                 \\
  $\vec{v}_i$               & velocity vector of star $i$                                      &                                                & \kms                               \\ \hline
  $\alpha_i$, $\delta_i$    & right ascension and declination of star $i$                      & fixed                                          & deg, deg                           \\
  $\vec{a}_i$               & astrometry vector $(\pi,\,\mu_\alpha,\,\mu_\delta)$ for star $i$ & observed                                       & $(\mathrm{mas},\,\masyr,\,\masyr)$ \\
  $\Cova$                   & noise covariance matrix of $\vec{a}$ for star $i$                & fixed                                          &                                    \\
  $v_{r,i}$                 & radial velocity                                                  & observed                                       & \kms                               \\
  $\sigma_{\mathrm{RV},i}$  & radial velocity error                                            & fixed                                          & \kms                              \\
  \hline
  \end{tabular}
  \caption{Summary of parameters and prior specifications for the model.} \label{tab:modelspec}
\end{table*}

\begin{figure}
  \centering
  \includegraphics[width=1\linewidth]{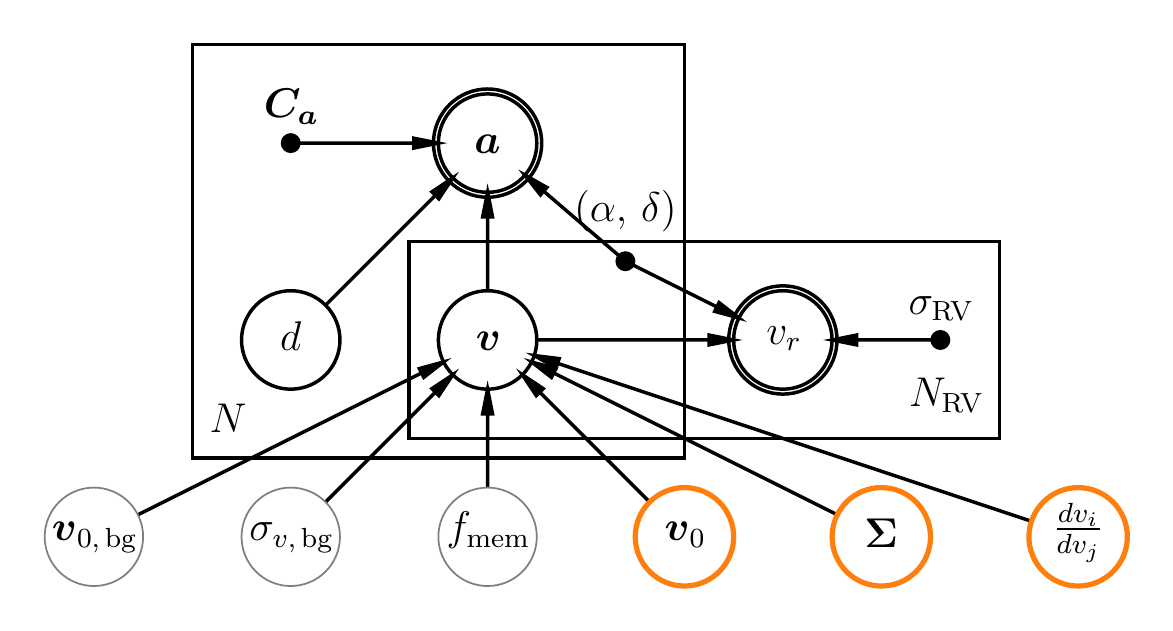}
  \caption{
    Probabilistic graphical model for cluster internal kinematics. The key
    parameters of interest -- mean velocity, velocity dispersion matrix and
    velocity gradient matrix -- are highlighted in thick orange circles,
    whereas the nuisance parameters -- mean velocity and (isotropic) velocity
    dispersion of the background, and the fraction of stars that are members
    -- are in thin gray circles. The observables (in double-lined circles)
    are the vectors $\vec{a} = (\parallax,\,\pmra,\,\pmdec)$ and radial
    velocities $v_r$ when available. Note that the uncertainty covariances
    between parallax and proper motions, \Cova, and uncertainties of radial
    velocities, $\sigma_{\mathrm{RV}}$ are taken into account. See
    \sectionname~\ref{sub:model} for details.
    }
  \label{fig:pgm}
\end{figure}

\subsection{Model} \label{sub:model}

Our model of the velocity field of cluster members includes linear velocity
gradients describing rotation and shear, and the full dispersion matrix as well
as contamination by kinematic outliers. Figure~\ref{fig:pgm} provides a
visual summary as a probabilistic graphical model. All parameters and their
priors (if any) are recorded in Table~\ref{tab:modelspec}.

Traditionally, the velocities of stars in a cluster are assumed to be the
same with a small, usually isotropic, dispersion. This assumption has been
utilized in mainly two different ways. If we have proper motions and radial
velocities (RVs) of the members, we can deduce the mean distance (parallax)
of the cluster (``moving cluster method''). On the other hand, if we have
parallaxes and proper motions, we can infer the mean and individual radial
velocities. The radial velocities derived in such a way are referred to as
``astrometric'' radial velocities in order to differentiate them from the
more common spectroscopic radial velocities from Doppler shifts
\citep{dravins1999-astrometric-rv}.
However, as discussed in the introduction, it is more interesting in the
\gaia-era to explore internal motions beyond this simple model for clues as
to the birth conditions or evolution of the cluster. We can do this by
combining precise astrometry with radial velocity measurements, which are
often only available for a portion of the astrometric sample.

The velocity of a member star $i$ in a cluster,
$\vec{v}_i$\footnote{All vectors are column vectors unless transposed.},
is assumed to be the sum of the mean velocity $\vec{v}_0$ of the cluster,
systematic peculiar velocity, and some scatter (dispersion), which is
expected to be small ($\lesssim$ few \kms).
Systematic internal motions such as rotation or shear
are captured by the linear velocity gradients $\mat{T}=T_{lk}=d v_l/d x_k$ to
first approximation \citep{lindegren2000}. The anti-symmetric part of
$\mat{T}$ describes the rigid body rotation, and the symmetric part describes
the extensional (contraction or expansion) and shear strain rates, which can
be diagonalized to examine the principal axes of shear:
\begin{equation}
  \frac{1}{2} (\mat{T} - \mat{T}^T) = 
    \begin{bmatrix}
      0 & -\omega_z & \omega_y \\
      \omega_z & 0 & -\omega_x \\
      -\omega_y & \omega_x & 0
    \end{bmatrix};\,\,
    \begin{split}
      \omega_x &= \frac{1}{2} (T_{zy} - T_{yz}) \\
      \omega_y &= \frac{1}{2} (T_{xz} - T_{zx}) \\
      \omega_z &= \frac{1}{2} (T_{yx} - T_{xy})
    \end{split}
\end{equation}
\begin{equation}
  \frac{1}{2} (\mat{T} + \mat{T}^T) = 
    \begin{bmatrix}
      w_4 & w_3 & w_2 \\
      w_3 & w_5 & w_1 \\
      w_2 & w_1 & 3\kappa - w_4 - w_5
    \end{bmatrix};\,\,
    \begin{split}
      w_1 &= \frac{1}{2} (T_{zy} + T_{yz}) \\
      w_2 &= \frac{1}{2} (T_{xz} + T_{zx}) \\
      w_3 &= \frac{1}{2} (T_{yx} + T_{xy}) \\
      w_4 &= T_{xx} \\
      w_5 &= T_{yy} \\
      \kappa &= \frac{1}{3} (T_{xx}+T_{yy}+T_{zz})
    \end{split}
\end{equation}
Here, $\omega_{\{x,\,y,\,z\}}$ is the rotation around each axis and we have
followed the notation of \citet{lindegren2000} to use $w_1, \ldots, w_5$ and
$\kappa$ for the components of the symmetric shear matrix.

It is well-known that there is a degeneracy between mean velocity
and the isotropic expansion/contraction component $\kappa$
when considering astrometry alone
\citep{blaauw1964-ScoCen,dravins1999-astrometric-rv}.
Generally, only 8 of 9 components of velocity gradient matrix $\mat{T}$ can
be determined from astrometry alone and there still exists one-dimensional
degeneracy due to lack of information on how radial velocities change.
By incorporating all radial velocities available for a subset of bright stars
in \gdrtwo, we can break the degeneracy and infer all nine components of $\mat{T}$ that
is most consistent with the data.

We make the velocity dispersion a general symmetric matrix, $\mat{\Sigma}$, in
order to test the assumption of isotropic dispersion in light of \gaia\ data.
Indeed, recent studies of young clusters including the Orion Nebula Cluster
have already reported anisotropic on-sky velocity dispersions with \gdrtwo\
\citep{kuhn2019,kim2019-onc,wright2019-Lagoon}. We decompose $\mat{\Sigma}$ into a
scale vector $\vec{\sigma}_v$ and correlation matrix $\mat{\Omega}$, such that
$\Sigma=\diag(\vec{\sigma}_v)\,\mat{\Omega}\,\diag(\vec{\sigma}_v)$.

As samples of cluster members are never perfect, it is important to account
for contamination by non-members (in terms of their velocity) in order to
make our inference robust to outliers \citep{hogg2010-fitting}.
We model the sample velocity
distribution as a mixture of two components: members which have velocities
drawn from the distribution described above and ``background'' non-members
which have a broad isotropic Gaussian distribution. This adds three more
parameters to the model, namely the fraction of stars that are members,
\fmem, and the mean and dispersion of the background velocities,
$\vec{v}_{0,\mathrm{bg}}$ and $\sigma_\mathrm{bg}$.
Putting this all together, we obtain:
\begin{equation}
    \vec{v}_i  \sim
    \begin{cases}
    \mathcal{N} \left( \vec{v}_0 + \mat{T} \cdot (\vec{b_i}-\vec{b_0}),\,\mat{\Sigma}\right)
      \equiv\mathcal{N}(\vec{v}_{0,\mathrm{cl}},\,\mat{\Sigma_\mathrm{cl}})\,\,\text{(cluster)} \\
    \mathcal{N} (\vec{v}_{0,\mathrm{bg}},\,\sigma_\mathrm{bg}^2 \mat{I})
      \equiv\mathcal{N}(\vec{v}_{0,\mathrm{bg}},\,\mat{\Sigma_\mathrm{bg}})
      \,\,\text{(background)}
    \end{cases}
    \label{eq:model}
\end{equation}
Here, $\vec{b}_i$ is the position vector to star $i$, whereas $\vec{b}_0$ is the (arbitrary) reference position vector where the velocity equals the mean velocity vector.
The observed proper motions of star $i$ are then projections of the velocity divided by distance, $\vec{v}_i/d$, in the right ascension (R.A.) and declination (Decl.) directions.

We pack the observables, parallax and proper motions, into
vector $\vec{a}_i$ for the $i$-th star as
\begin{equation}
  \vec{a}_i = 
\begin{bmatrix}
  \pi_i \\
  \mu_{\alpha,i} \\
  \mu_{\delta,i}
\end{bmatrix}.
\end{equation}
Then, the mean model of $\vec{a}_i$, which we note with $\bar{\vec{a}}_i$
is related to velocity and distance, $\vec{v}_i$ and $d_i$, as
\begin{equation}
  \bar{\vec{a}}_i (d_i, \vec{v}_i) = 
\begin{bmatrix}
  1/d_i \\
  \vec{p}^T_i \vec{v}_i / d_i \\
  \vec{q}^T_i \vec{v}_i / d_i
\end{bmatrix}
\end{equation}
where $\vec{p}_i$ and $\vec{q}_i$ are unit vectors in R.A. and Decl. direction
at the position of star $i$. We assume a Gaussian noise model for \gaia,
i.e., $\vec{a}_i \sim \mathcal{N}(\bar{\vec{a}}_i,\,\mat{C}_{\vec{a},i})$
\citep{hogg2018-gaia-likelihood}.
Since the noise model and the velocity dispersion are both Gaussian,
we can exploit the self-conjugacy of Gaussian distributions and
marginalize over $\vec{v}_i$ analytically \citep{lindegren2000}.
Then, the likelihood of $\vec{a}_i$ is directly related to the hierarchical
parameters, $\{\vec{v}_0, \mat{T}, \mat{\Sigma}\}$ or
$\{\vec{v}_{0,\mathrm{bg}}, \sigma_\mathrm{bg}\}$ depending on which
mixture component we are concerned with as
\begin{equation}
  \vec{a}_i \sim \mathcal{N}(\bar{\vec{a}}_i(d_i, \vec{v}_{0,\mathrm{cl/bg}}),\,
    \mat{D}_i(d_i, \mat{\Sigma}_\mathrm{cl/bg}))
\end{equation}
where the modified covariance matrix $\mat{D}_i$ is the sum of the
observational covariance $\mat{C}_{\vec{a},i}$ given in \gdrtwo\
and the projected velocity dispersion converted to proper motion dispersion at
the star's position:
\begin{equation}
  \mat{D}_i (d_i,\mat{\Sigma})= \mat{C}_{\vec{a},i} + \frac{1}{d_i^2}
  \begin{bmatrix}
  0 & \begin{matrix} 0 & 0 \end{matrix}\\
  \begin{matrix} 0 \\ 0 \end{matrix} & \mat{M}^T_i \mat{\Sigma} \mat{M}_i
  \end{bmatrix}.
\end{equation}
Here, $\mat{M}_i = [\vec{p}_i,\,\vec{q}_i]$.

We combine radial velocity measurements to the likelihood when available
(\figname~\ref{fig:pgm}).
It is straightforward to extend this to radial velocities:
\begin{equation}
  \begin{split}
    &\bar{v}_{r,i}(\vec{v}_i) = \vec{r}_i^T\vec{v}_i \\
    &v_{r,i} \sim \mathcal{N}(\bar{v}_{r,i}(\vec{v}_{0,\mathrm{cl/bg}}),
    \,\sigma_\mathrm{RV}^2 + \vec{r}_i^T \mat{\Sigma_{\mathrm{cl/bg}}} \vec{r}_i)
  \end{split}
\end{equation}
where $\vec{r}_i$ is the unit vector in radial direction at the position of star $i$.
Note that $\{\vec{p}_i, \vec{q}_i, \vec{r}_i\}$ forms an orthonormal basis that depends
on the R.A. and Decl. of star $i$.

Now we can write down the full likelihood for each component of the mixture,
\begin{equation}
  \ln \mathcal{L_{\mathrm{cl/bg}}} = \sum_{i} \ln p(\vec{a}_i \given \vec{\theta}_\mathrm{cl/bg}) +
    \sum_{i \in I_\mathrm{RV}} \ln p(v_{r,i} \given \vec{\theta}_\mathrm{cl/bg})
\end{equation}
where $I_\mathrm{RV}$ denotes the index set of stars with RVs and $\vec{\theta}$
is the vector containing all parameters of each mixture component.
Finally, the combined likelihood of the mixture model introduces
one more parameter $\fmem$, the fraction of stars in the cluster component:
\begin{equation}
  \centering
  \mathcal{L} = \fmem \mathcal{L}_\mathrm{cl}
    + (1-\fmem) \mathcal{L}_\mathrm{bg}.
\end{equation}

We specified broad Gaussian priors for the mean velocities of the cluster and
the background, the velocity dispersion of the background (which is
constrained to be positive), and the velocity gradients. A uniform density
prior was assumed for the fraction of stars that are cluster members although
we know the contamination fraction is quite small. For the velocity
dispersion matrix of the cluster, we specified half-Cauchy distributions with
the scale parameter $\gamma=2.5$ for the scale vector,
and LKJ prior \citep{lkj2009-lkj-prior} with $\eta = 2$ for the correlation matrix
-- see Table~\ref{tab:modelspec} for quantitative details.

We stress that the perspective effect is fully taken into account by projecting
velocities $\vec{v}_i$ at each star's location on the celestial sphere
$(\alpha_i,\,\delta_i)$ using the basis
$\{\vec{p}_i,\,\vec{q}_i,\,\vec{r}_i\}$ to forward-model the parallaxes and
proper motions.
In fact, the perspective effect of all components of the velocity, that is not only
the mean velocity $\vec{v}_0$ but also the velocity gradient $\mat{T}$ and
anisotropic dispersion $\mat{\Sigma}$, are naturally handled \emph{exactly}
without the first-order approximation \citep[e.g.,][]{kuhn2019,vanLeeuwen2009}
regardless of how large the object is on the sky
(under the assumption that uncertainties of $(\alpha_i,\,\delta_i)$ are negligible).
Of course, the dominant perspective effect is due to the mean velocity $\vec{v}_0$
and generally it can be both expansion/contraction or rotation-like patterns
in the projected position-velocity space, $(\alpha,\,\delta)$ vs. $(v_\alpha,\,v_\delta)$
depending on the mean velocity and the position of the cluster on the celestial sphere.
We expand on the perspective effect of the mean velocity in
Appendix~\ref{appendix:perspective-effect}, contrasting our method with
the first-order correction and providing validation that they do not
affect the inference of velocity gradient $\mat{T}$.

\subsection{Implementation \& Validation} \label{sub:validation}

\begin{figure*}
	\centering
	\includegraphics[width=0.75\linewidth]{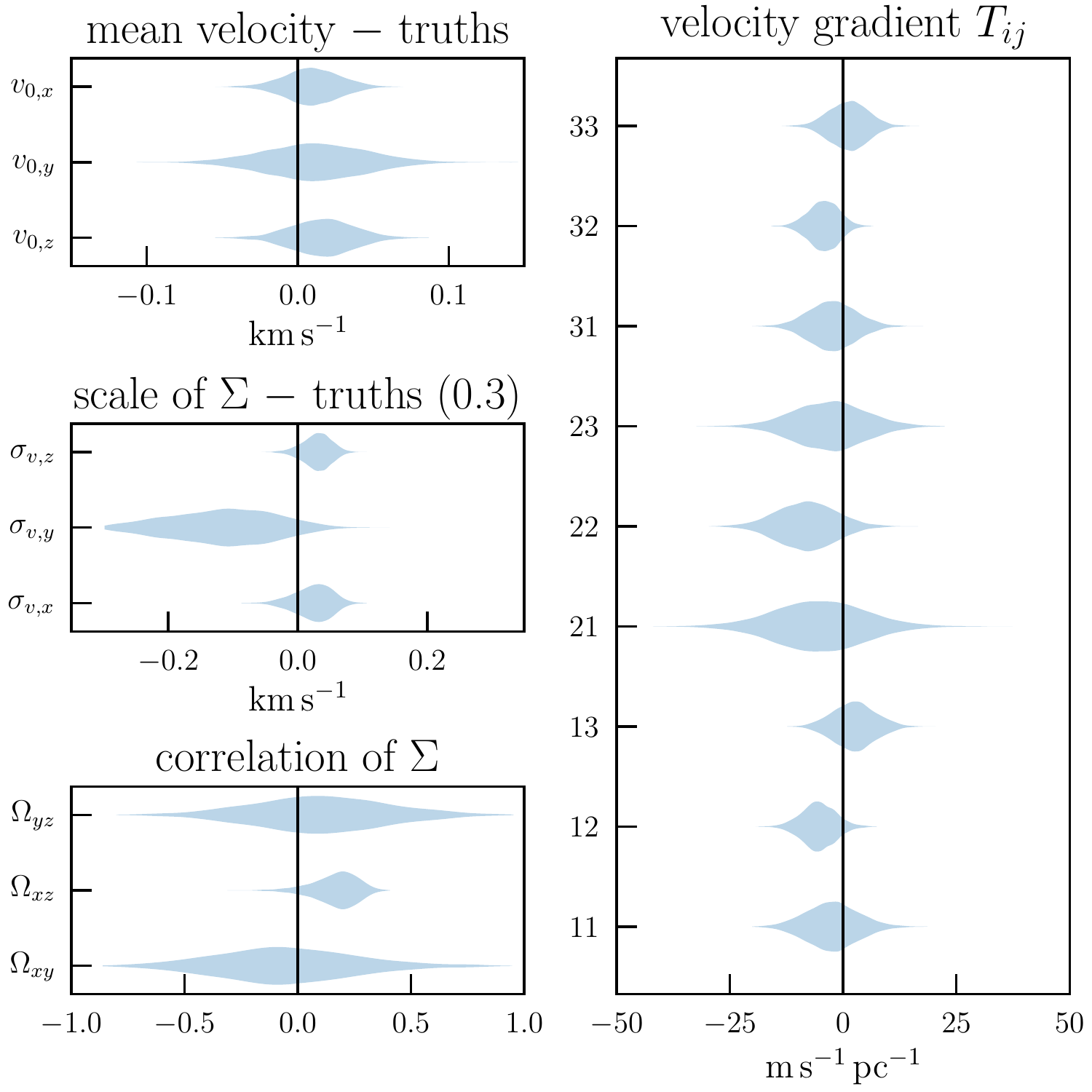}
	\caption{
    Model validation with mock data generated with \gdrtwo\ quality analogous
    to real data for the Hyades cluster. Each violin plot shows the posterior
    pdf of the parameter labeled on the vertical axis. We subtract the true
    value set to generate the mock data, thus ideally distributions should
    include $0$ (vertical line). The mock data in this case was generated
    with an isotropic velocity dispersion and no velocity gradient. All
    parameters here are in the ICRS coordinate system.
	}
	\label{fig:validation}
\end{figure*}

We implement the model in Stan\footnote{\url{https://mc-stan.org}}, a
probabilistic modelling software \citep{carpenter2017} using the PyStan
interface\footnote{\url{https://mc-stan.org/users/interfaces/pystan}}.
Once the generative model is specified in its own Stan language, we can
either optimize using e.g., (quasi)-Newton's methods to get a point estimate
of parameters or sample the joint posterior distribution using the no-U-turn
sampling (NUTS) algorithm \citep{nuts}, an extension to Hamiltonian Monte
Carlo (HMC) that eliminates fine-tuning of sampling parameters which can have
significant effect on its sampling efficiency. HMC requires derivatives of
the target density function with respect to the parameters but once tuned,
can sample high-dimensional parameter spaces efficiently. Since our model is
analytically differentiable with $\mathcal{O}(N)$ parameters where $N \approx
500-1000$ (number of stars), Stan and its NUTS sampling are well-suited.
  
In all following inferences, we sample the model parameters using NUTS with 4
chains and 2000 iterations. Discarding the first half as ``warm-up'' produces
$4\times1000=4000$ samples in total. We check the Gelman-Rubin statistic
$\hat R$ of the posterior draws to ensure that the chains have converged
\citep{gelman-rubin1992,vehtari-improved-rhat2019}.

We test our implementation using mock data of \gdrtwo\ quality generated
according to the model (\sectionname~\ref{sub:model}), assuming a fiducial
set of parameters for the Hyades cluster. The mock data was generated using
the exact \gdrtwo\ sky positions and uncertainties of the actual Hyades data
which we later model (the `cl' sample described in
\sectionname~\ref{sub:data}). We assumed the mean velocity and isotropic
velocity dispersion similar to the actual values, but zero velocity gradient:
\begin{equation}
  \begin{split}
    \vec v_0 &= (-6.32,\,45.24,\,5.30)~\kms \\
    \mat{\Sigma} &= 0.3 \mat{I}~(\kms)^2 \\
    \mat{T} &= 0.
  \end{split}
\end{equation}
This is the simplest null case in which the velocity dispersion is isotropic and
there is no rotation or shear. We may contrast this with our fits to the
actual data to gauge the significance. We added 10\% contamination from a
broad background model.

The two component mixture model correctly labels cluster members and
background contamination. \figname~\ref{fig:validation} shows the posterior
distribution of model parameters minus their true values. Ideally, the
distributions should include zero (vertical lines) meaning that we recover
the true parameters put in.
We find that all parameters are well recovered.
The velocity dispersion along $y$-axis is
biased towards a smaller value by $0.12$~\kms. There are two factors that may
bias the velocity dispersion to be smaller than it is. One is if the velocity
errors are too large making the internal dispersion unresolved. Another is
when there is lack of information on velocity in a given direction and the
prior (peaked at $0$) drives the posterior distribution. The primary reason
here is likely the first as the median velocity error ($0.43$~\kms) is
slightly larger the assumed dispersion. This is also why 
parameters involving the $y$-axis have a larger uncertainty compared to the others.
Nonetheless, by incorporating partial RVs we can correctly infer null
velocity gradient within $\approx 10$~\mspc.
Determining all nine components of $\mat{T}$ is only possible when
including radial velocities available for a subset of stars,
as we discussed in \sectionname~\ref{sub:model}.

\section{Application to the Hyades cluster}
\label{sec:application}

We apply the method to the \gdrtwo\ data of the Hyades cluster and its tails
in order to examine the internal motions in light of the improved data
quality.

\subsection{Data}
\label{sub:data}

\begin{table*}
  \begin{tabular}{llrrrl}
    \toprule
    Designation &  R.A. & Decl.  &  $\langle P_{\mathrm{mem},i} \rangle$ & fit group \\
    \midrule
     Gaia DR2 49520255665123328 &  64.874609 &  21.753716 &   0.997003 &        cl \\
     Gaia DR2 49729231594420096 &  60.203783 &  18.193881 &   0.996497 &        cl \\
     Gaia DR2 51383893515451392 &  59.806965 &  20.428049 &   0.998720 &        cl \\
    Gaia DR2 145373377272257664 &  66.061268 &  21.736049 &   0.999570 &        cl \\
    Gaia DR2 145391484855481344 &  67.003711 &  21.619722 &   0.984320 &        cl \\
    \bottomrule
  \end{tabular}
  \caption{
    The merged Hyades sample used for kinematic modelling.
    We present the first five rows and make the full table available online as supplementary
    material.
    In the full table, we add parallaxes, proper motions, radial velocities,
    \bprp\ colors, and $G$ magnitudes from \gdrtwo\
    as well as boolean columns indicating whether the source was included
    the membership list of \citet{gaiacollab-dr2-hrd}, \citet{meingast2019} or
    \citet{roser2019}.
    In addition to the available data, we provide the mean membership probability from
    our kinematic modelling.
    \label{tab:fullsample}
  }
\end{table*}

\begin{figure*}
  \centering
  \includegraphics[width=0.95\linewidth]{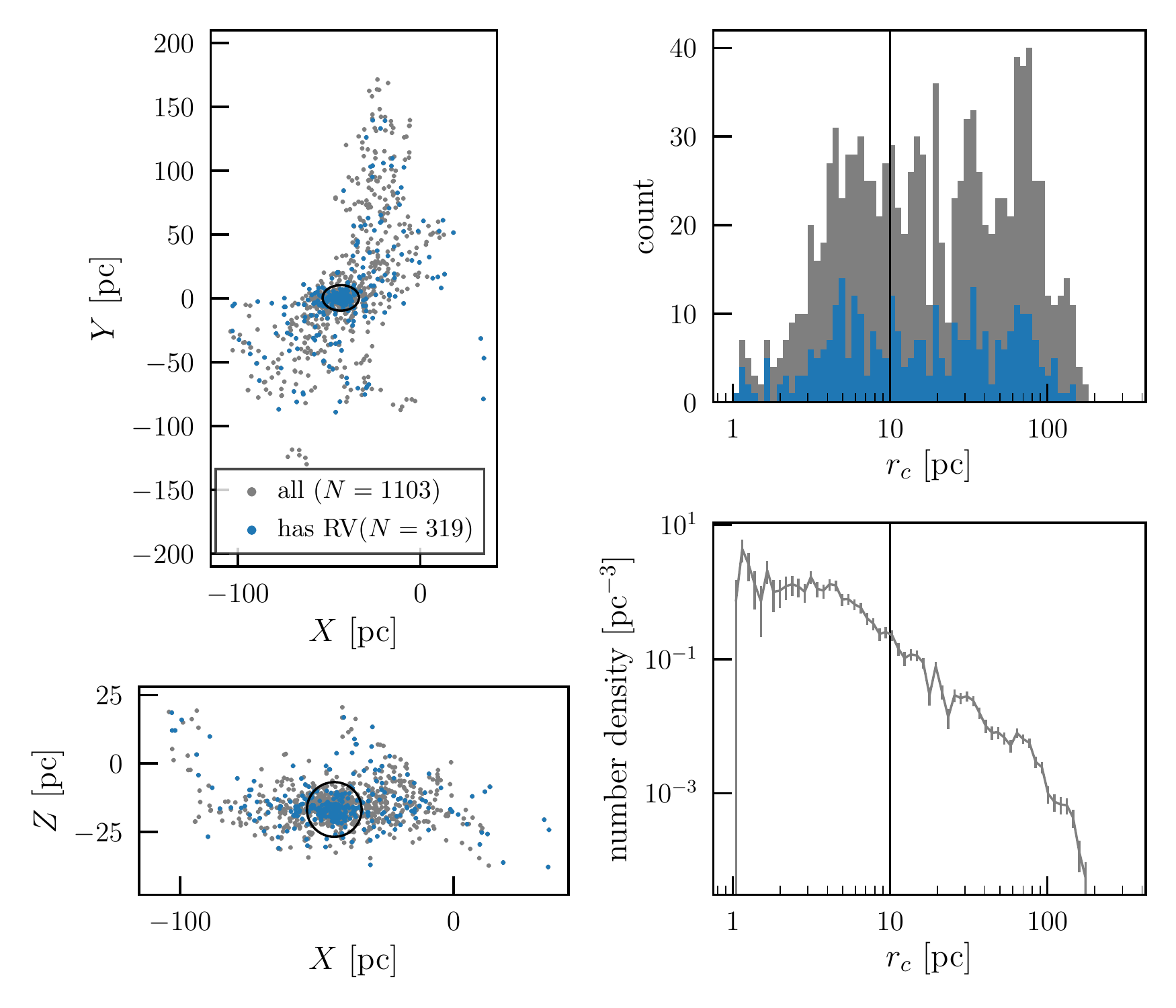}
  \caption{
    Distribution of the Hyades cluster sources in the merged sample.
    Left panels: Distribution of the Hyades cluster sources 
    in Galactic coordinates ($X,Y,Z$) centered on the Sun. The
    Galactic centre is towards the right ($+X$) and the Galactic rotation is
    up ($+Y$). Sources with radial velocities available in \gdrtwo\ are
    highlighted in blue. The black circle around the cluster centre marks
    $10$~pc radius, dividing the `cl' ($r_c<10$~pc) and `tails' ($r_c>10$~pc)
    sample which we model separately. Right panels: Histogram and number
    density of Hyades sources as a function of cluster-centric distance
    $r_c$. Details are discussed in \sectionname~\ref{sub:data}.
    }
  \label{fig:hyades_dist}
\end{figure*}

We use the cluster member selection of \citet{gaiacollab-dr2-hrd} as our base
sample and merge this with two different samples of tidal tails by
\citet{meingast2019} and \citet{roser2019}. The motivation for considering
tidal tails is that any signature of a non-zero linear velocity field is
larger at larger distances from the cluster center. In particular, shear due
to Galactic tides is stronger for stars beyond the tidal radius that are
farther away from the cluster potential. Both selections were made to find
stars that have similar velocity with an assumed Hyades cluster mean velocity
within $200$~\pc\ distance from the Sun with some density threshold to reduce
contamination by unrelated field stars having coincidentally similar
velocities. However, \citet{roser2019} made the selection from all DR2
astrometric sources (with quality cuts to clean unreliable measurements)
while \citet{meingast2019} only selected from bright sources with their RVs
measured. We exclude sources classified as ``other'' by \citet{roser2019} that are
significantly more spatially offset from the rest, as
they are most likely not part of the Hyades cluster or its tails. In summary,
there are 92 sources added from \citet{meingast2019} and 568 sources added
from \citet{roser2019} to the base sample.

The merged sample consists of 1103 sources and is presented in Table~\ref{tab:fullsample}.
\figname~\ref{fig:hyades_dist} shows the distribution of the sample in
Galactic coordinates. Since the dynamics of stars differs within and beyond
the tidal radius, we divide the sample into two based on cluster-centric
distances. First, we determine the cluster centre iteratively as the mean
position of stars within a radius cut using the entire sample. We start with
the mean position
of all stars and select stars within the chosen radius cut. We determine a
new centre as the mean of those stars and repeat until the stars we select to
be within the radius cut converges. The radius cut should be large enough so
that the mean position is not dominated by statistical fluctuations of small
number of stars, but small enough so that the increasing contamination of
kinematic outliers do not affect the mean position. We choose $10$~pc as our
radius cut, which is comparable to the tidal radius of the cluster. After 10
iterations, we determine the centre as the mean position of 400 stars within
10 pc:
$b_c = (17.154,\,41.289,\,13.691)~\pc$ (ICRS) or $(-43.629,\,0.336,\,-16.820)$.
Note that our goal here is not to determine the centre of mass of the cluster
very accurately, which requires assigning mass to each star, but to come up
with a reasonable reference position for the cluster centre in order to study
how the kinematics of stars change with cluster-centric distance.

With the cluster centre determined, we divide the sample into two: 400 stars
within $10$~pc which we call `cl' and 703 stars beyond $10$~pc which we call
`tails'.
\gdrtwo\ RVs are available for 127 and 192 sources in cl and tails sample respectively.
The distribution of these sources are highlighted in \figname~\ref{fig:hyades_dist}
as blue circles.
They are spread around in all spatial dimensions providing anchor points
to break the degeneracy between perspective effect and velocity gradients.
The median velocity uncertainty in R.A. and Decl. direction
is $0.093$ and $0.043$~\kms\ while the median radial velocity error
is 0.43~\kms.

Out of concern that the velocity gradient $\mat{T}$ may be washed out by
some small shift in the mean velocity $\vec{v}_0$ if left as a free parameter
when modelling the tails, we fix the mean velocity to
that inferred from modelling the cluster (mean of the posterior samples).
We have also compared the results with when the mean velocity is still left a free
parameter for tails, and found that the mean velocity inferred from
tails is statistically the same as the cluster and that there are no
significant discrepancies in other parameters.

\subsection{Results}

\begin{table*}
\begin{tabular}{lrrrrrrrr}
\toprule
{} & \multicolumn{4}{c}{cl} & \multicolumn{4}{c}{tails} \\
{} &    mean &      sd & hpd 3\% & hpd 97\% &    mean &     sd & hpd 3\% & hpd 97\% \\
\midrule
\fmem                  &   0.953 &   0.013 &   0.929 &    0.976 &   0.870 &  0.014 &   0.844 &    0.897 \\
$v_{0,x}$ (ICRS)       &  -6.086 &   0.029 &  -6.144 &   -6.036 &         &        &         &          \\
$v_{0,y}$ (ICRS)       &  45.629 &   0.050 &  45.539 &   45.724 &         &        &         &          \\
$v_{0,z}$ (ICRS)       &   5.518 &   0.025 &   5.471 &    5.563 &         &        &         &          \\
$\sigma_{x}$           &   0.442 &   0.070 &   0.304 &    0.561 &   0.807 &  0.050 &   0.717 &    0.905 \\
$\sigma_{y}$           &   0.383 &   0.017 &   0.352 &    0.414 &   0.515 &  0.035 &   0.452 &    0.581 \\
$\sigma_{z}$           &   0.371 &   0.056 &   0.270 &    0.470 &   0.389 &  0.017 &   0.359 &    0.421 \\
$\Omega_{xy}$          &  -0.146 &   0.370 &  -0.837 &    0.502 &   0.487 &  0.086 &   0.324 &    0.642 \\
$\Omega_{xz}$          &  -0.015 &   0.297 &  -0.561 &    0.536 &   0.197 &  0.061 &   0.081 &    0.312 \\
$\Omega_{yz}$          &  -0.165 &   0.171 &  -0.455 &    0.161 &   0.122 &  0.066 &   0.002 &    0.246 \\
$\omega_x$             &   3.270 &   5.513 &  -6.570 &   14.245 &   5.268 &  2.241 &   1.129 &    9.633 \\
$\omega_y$             &   2.236 &   9.779 & -15.803 &   21.091 &   3.613 &  3.256 &  -2.587 &    9.714 \\
$\omega_z$             &  -4.440 &   8.713 & -20.170 &   12.407 &  -6.476 &  1.153 &  -8.721 &   -4.342 \\
$w_1$                  &   1.447 &   5.461 &  -9.050 &   11.613 &  -2.498 &  2.219 &  -6.631 &    1.785 \\
$w_2$                  &  -6.589 &  10.074 & -25.598 &   12.016 &  -2.156 &  3.368 &  -8.079 &    4.556 \\
$w_3$                  &   1.656 &   8.696 & -15.204 &   17.449 &  16.897 &  0.916 &  15.250 &   18.724 \\
$w_4$                  & -11.191 &  15.598 & -41.540 &   18.310 &   4.274 &  2.204 &   0.172 &    8.513 \\
$w_5$                  &  10.643 &   6.322 &  -0.938 &   23.013 &   0.712 &  1.223 &  -1.529 &    3.037 \\
$\kappa$               &  -6.500 &   6.417 & -18.385 &    5.434 &  -5.966 &  1.212 &  -8.163 &   -3.650 \\
$v_{\rm{bg},x}$ (ICRS) &  -5.948 &   0.536 &  -6.927 &   -4.878 &  -5.440 &  1.225 &  -7.667 &   -3.073 \\
$v_{\rm{bg},y}$ (ICRS) &  46.301 &   0.637 &  45.143 &   47.532 &  37.817 &  1.497 &  35.113 &   40.740 \\
$v_{\rm{bg},z}$ (ICRS) &   5.421 &   0.524 &   4.485 &    6.441 &   2.217 &  1.244 &   0.002 &    4.752 \\
$\sigma_{\rm bg}$      &   2.035 &   0.267 &   1.584 &    2.547 &  11.119 &  0.572 &  10.090 &   12.243 \\
\bottomrule
\end{tabular}

  \caption{
    Statistical summary of posterior distribution of parameters for cl and
    tails fits. For each parameter, we quote mean, standard deviation (sd),
    and (3, 97)\% highest posterior density interval (hpd 3\% and hpd 97\%).
    Note that for the tails ($r_c>10$~\pc) fit, $v_0$ is fixed to the
    mean of cl fit. Values are in the Galactic frame unless otherwise noted as
    ICRS.
  } \label{tab:result}
\end{table*}

\begin{figure*}
  \centering
  \includegraphics[width=0.95\linewidth]{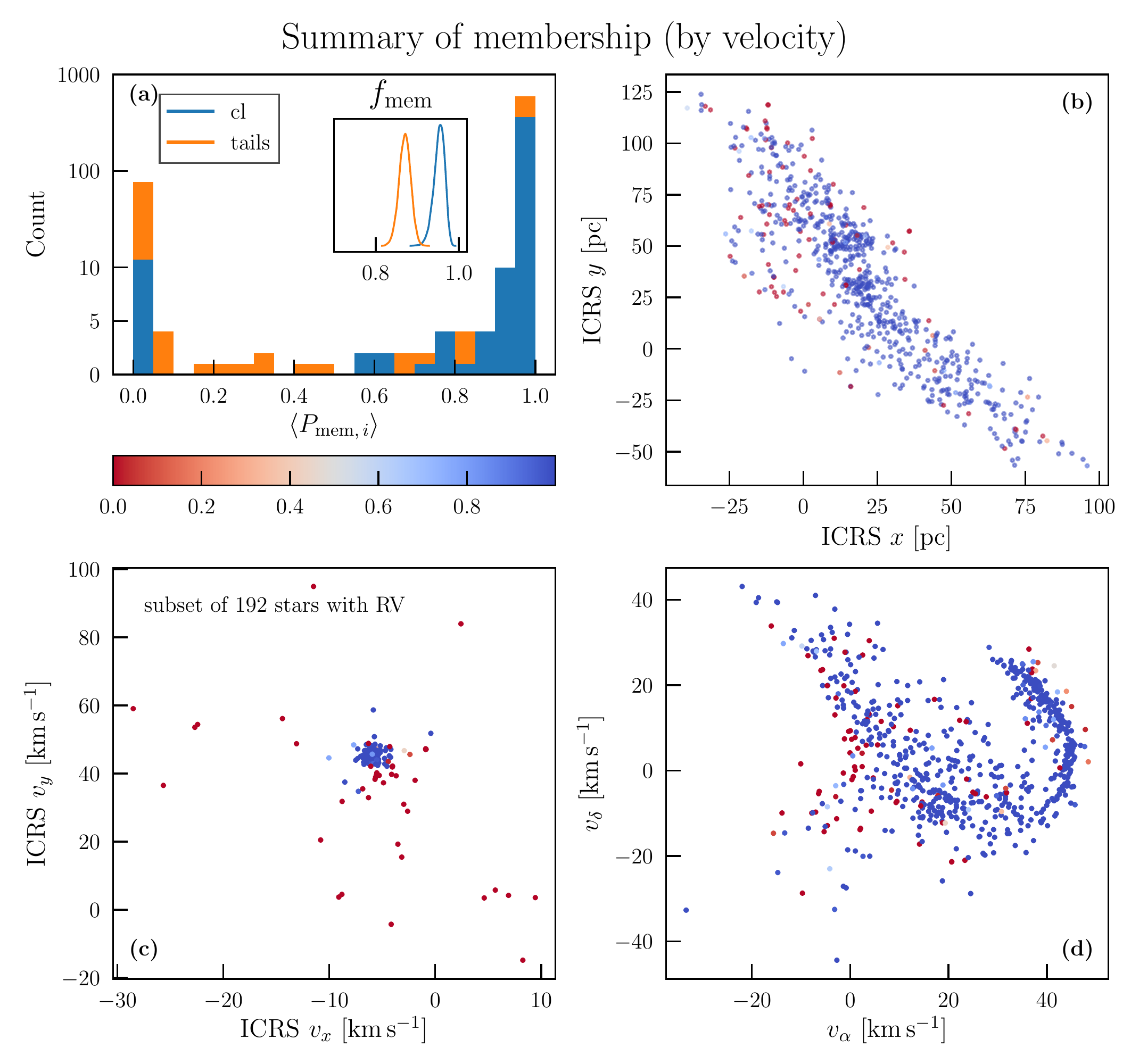} 
  \caption{Summary of membership by velocity from the mixture-model fitting.
    Panel (a) shows the distribution of mean membership probability of
    individual stars for cl and tails fits (blue and orange, same as in the
    other figures), while the inset shows the posterior pdfs of the fraction
    of stars which are members, $f_\mathrm{mem}$. The distribution of
    membership probabilities is highly bimodal, i.e., there is little ambiguity
    in the cluster membership by velocity from these (already filtered)
    data.
    The mean membership fraction of stars for cl and tails fits are $0.95$
    and $0.87$ respectively (Table~\ref{tab:result}).
    We show the distribution of stars for the tails
    coloured by their mean membership probability in three different
    two-dimensional projections of the data: in cartesian ICRS coordinates (b), in
    cartesian ICRS velocities (c); for a subset of stars with RVs (although we
    model all stars with astrometry) and in on-sky velocities (d). The
    membership here is defined only by the velocity vector.}
  \label{fig:membership}
\end{figure*}

\begin{figure*}
  \centering
  \includegraphics[width=0.95\linewidth]{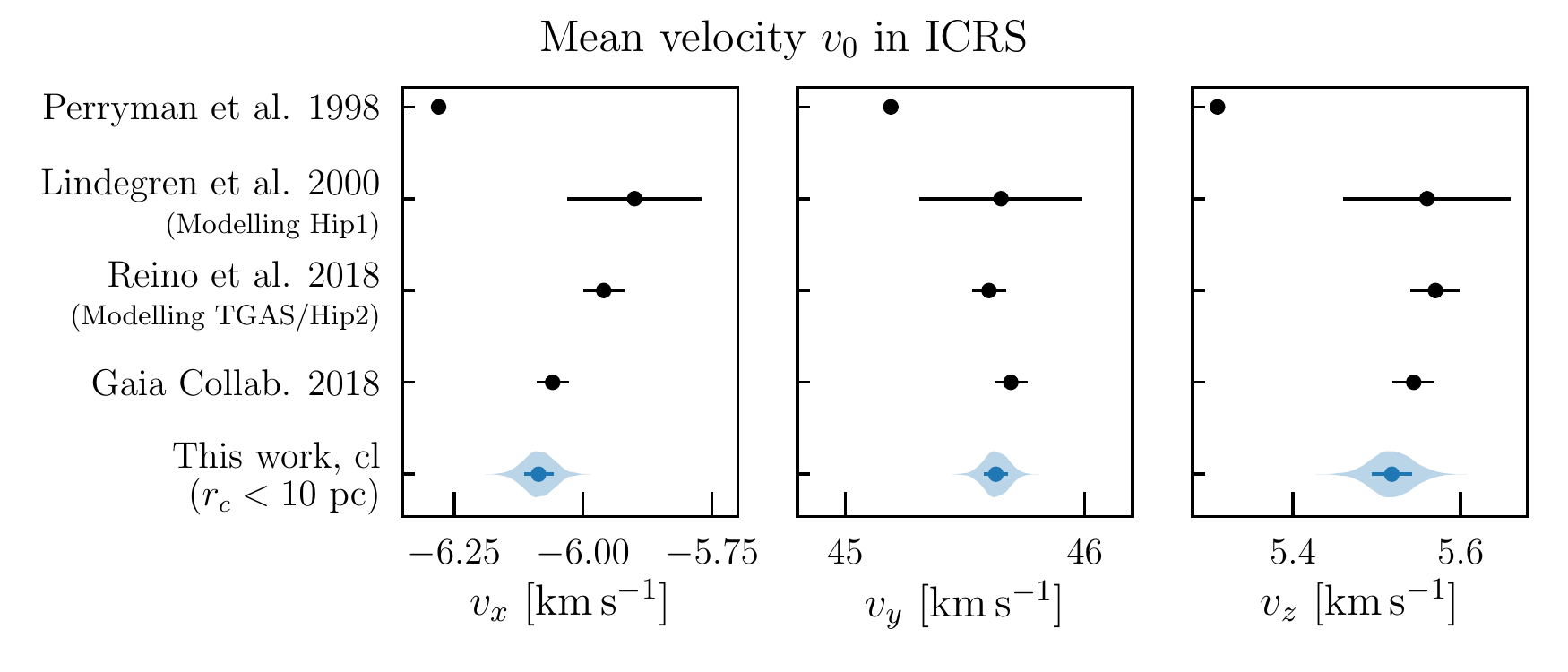} 
  \caption{
    Mean velocity $v_0$ inferred from the cl fit (see
    \sectionname~\ref{sub:data}), compared with previous studies. For this
    work, the shaded region (violin plot) visualizes the distribution of
    posterior probability density while the marker and error bars correspond
    to the median and (16, 84)\% confidence interval (Gaussian $1\sigma$) in
    accordance with the other works.
  }
  \label{fig:mean_velocity_comparison}
\end{figure*}

\begin{figure*}
  \centering
  \includegraphics[width=0.95\linewidth]{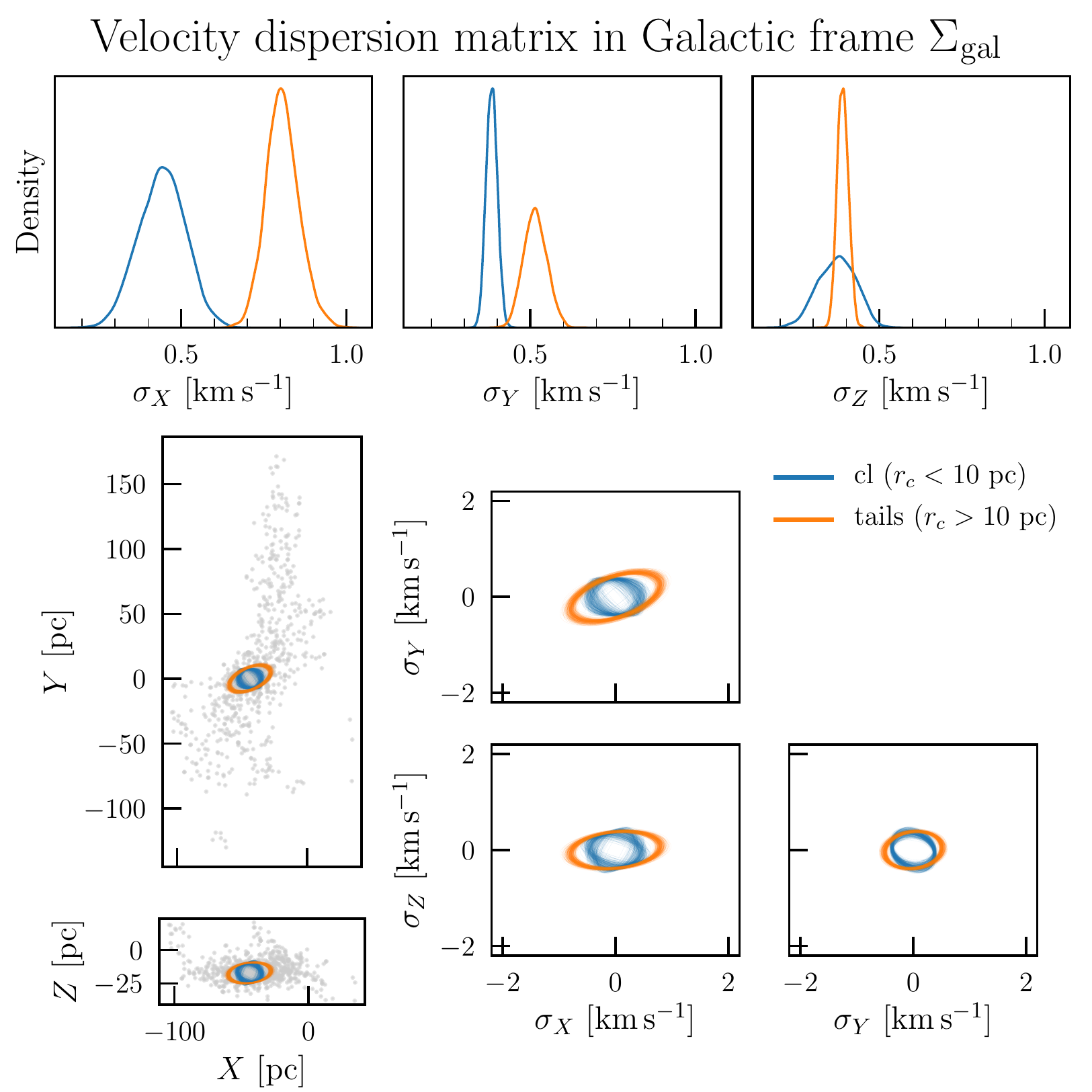} 
  \caption{
    Velocity dispersion matrix $\Sigma_{\rm gal}$ in Galactic coordinates.
    The top three panels show the posterior pdf of the scale in $X$, $Y$ and
    $Z$ directions (with $+X$ in the direction towards the Galactic centre
    and $+Y$ in the direction of Galactic rotation). The bottom right corner
    plots visualize the $1\sigma$ dispersion ellipses projected on $2D$
    planes from 400 (10\%) randomly selected posterior samples. The same
    dispersion ellipses are plotted on top of the distribution of stars in
    Galactic $(X,Y)$ and $(X,Z)$ planes on the left, centered at the cluster
    centre.
    }
  \label{fig:Sigma_gal}
\end{figure*}

\begin{figure*}
  \centering
  \includegraphics[width=0.85\linewidth]{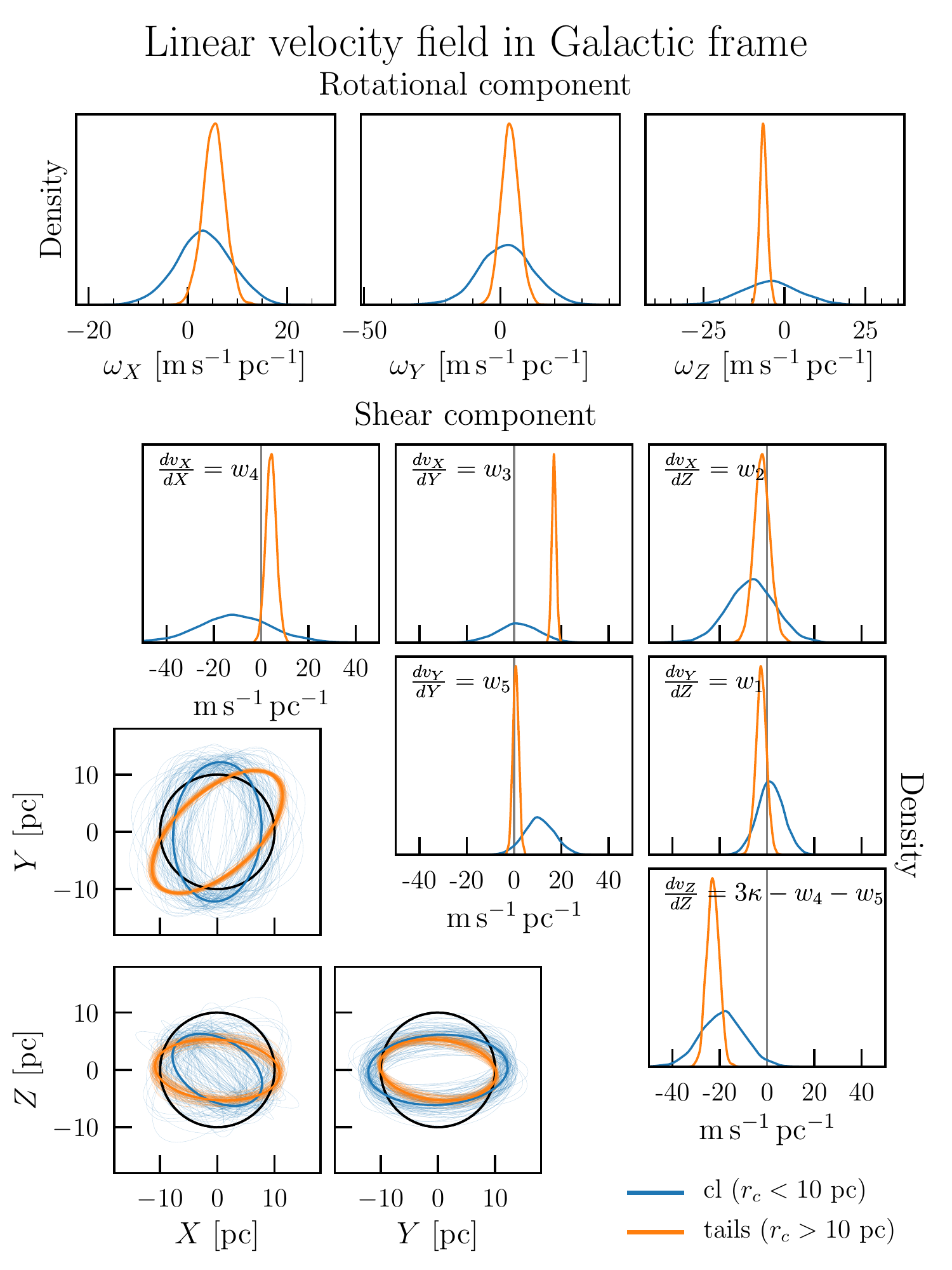} 
  \caption{
    Decomposed linear velocity field inferred from cl and tails fits.
    The top row shows the rotation in the Galactic frame. The posterior pdfs for each
    component of the shear component of the linear velocity field (symmetric
    part of the tensor $T_{ij}$) is shown in the upper corner plot. On the
    bottom left corner plot, we visualize the shear by how it would deform a
    circle with radius of 10~\pc\ in 50~Myr. The thicker line shows the mean
    of all posterior samples while each thin ellipse is from a hundred (25\%)
    posterior samples randomly selected. We find negligible rotation in both
    cl and tails fits. On the other hand, there is significant positive
    shear in specific directions (in which ellipses are elongated) and
    negative (compressive) shear along the Galactic $Z$ direction.
  }
  \label{fig:T_gal}
\end{figure*}

We present and compare the results of cl and tails fits in the order of
membership (\fmem), mean velocity ($v_0$), velocity dispersion matrix
($\Sigma$) and linear velocity gradient ($dv_i/dv_j$), each summarized in
\figname{s}~\ref{fig:membership}, \ref{fig:mean_velocity_comparison},
\ref{fig:Sigma_gal} and \ref{fig:T_gal}. A statistical summary of the
posterior samples is provided in Table~\ref{tab:result}.

\figname~\ref{fig:membership} summarizes the membership from the simultaneous
modelling of all parameters. Generally, both cl and tails have
low contamination fraction and the kinematic outliers
are well-separated from the members as shown in \figname~\ref{fig:membership} (a).
The rest of the panels in \figname~\ref{fig:membership} show the distinction
between members and non-members by the velocity mixture model in various projections
of the data for tails, where each source is coloured by
its mean posterior membership probability $\langle P_{\mathrm{mem},\,i}\rangle$ indicated
in the colour bar:
positional (b), cartesian velocity (c) and projected velocity (d) space.
Naturally the distinction is most clear cut in cartesian velocity space (c)
but note that while we can only put stars with RVs on this diagram, the rest
of the data without RVs are consistently and simultaneously modelled and
shown in (b) and (d). In projected velocities, stars with the same velocity
can exhibit a non-trivial trajectory due to changing perspective (d). Most
importantly, this nuisance, i.e., the existence of kinematic outliers, is
marginalized out in our inference of internal kinematics, making the results
robust to contamination in member selection.
We make the (kinematic) membership probability from our analysis
available (Table~\ref{tab:fullsample}), as this may be useful for other applications.

\figname~\ref{fig:mean_velocity_comparison} shows the inferred mean velocity
$v_0$ of the cluster in ICRS coordinates when modelling the cluster
proper (cl) in comparison with a number of previous studies.
We find our estimate for the mean velocity to be consistent with
\citet{gaiacollab-dr2-hrd}, which is also from the \gdrtwo\ data, as well as
previous studies modelling the \tgas\ and \hipparcos\
data~\citep{reino2018-tgas-hyades,lindegren2000}.
The main difference is that the uncertainties are smaller, thanks to better
quality and larger size of \gdrtwo\ data.

In previous kinematic modelling of the Hyades, the velocity dispersion was
assumed to be isotropic. Moreover, the larger noise in proper motions and
parallaxes and the lack of RVs meant that the small internal dispersion is
only marginally resolved. With \gdrtwo\ and a more flexible model, we find
that the velocity dispersion is indeed mildly anisotropic for the cluster
(cl ($r_c<10$), \figname~\ref{fig:Sigma_gal}).
However, the velocity ellipsoid of the tails is strongly anisotropic and
elongated in the Galactic radial direction, while remaining unchanged in Galactic vertical direction.
Hints of this velocity dispersion anisotropy
can be already seen with \tgas\ data
\citep[][\figname~14]{reino2018-tgas-hyades}; it is most clear from the
bottom row of their figure where the velocities are calculated using the
kinematically-improved parallaxes from their kinematic modelling and is
largely consistent with what we find, although modelling is required to
deconvolve the noise and covariance from the apparent dispersion.

In reality, the velocity dispersion likely changes with the
cluster-centric distance. Generally, the velocity dispersion decreases with
radius, but may deviate from the expectation of isolated bound cluster
starting at $\approx$ half tidal radius. This is because, under the influence
of the tidal field, a population of stars that are energetically unbound yet
still within the tidal radius (``potential escapers'') may increasingly
dominate the kinematics of a cluster \citep[e.g.,][]{Ba01,kupper2010}.
Figure 12 of \citet{Ba01} indicates that the fraction of potential escapers may be $\approx 20 \%$ for the Hyades.
Beyond the tidal
radius, the velocity dispersion of the tidal debris increases with radius.
While we do not model the velocity dispersion as a function of
cluster-centric radius, the increase of the inferred velocity dispersion in
the Galactic radial and azimuthal directions for the tails compared to the
cluster is consistent with this expectation \citep[see also][]{meingast2019,ernst2011}.
On the other hand, the velocity dispersion
in the Galactic vertical direction remains almost unchanged.

Finally, the posterior probability distribution of the linear velocity field
parameters are presented in \figname~\ref{fig:T_gal}. We transform the
velocity gradient tensor to more physically interpretable components, namely
rotation and shear. We do not find any significant rotation in both fits.
For the cluster (cl, $r_c<10$~\pc), there is no net expansion or contraction
but there is $\approx 2\sigma$ level shear signals that the cluster
is being stretched in the Galactic rotational direction ($Y$ axis) and
compressed in the Galactic vertical direction ($Z$ axis). A similar shear
field is much more significantly detected in tidal tails and its direction is
more well-defined.

\subsection{Discussion}

\subsubsection{Effects of \gaia\ Frame Rotation for Bright Sources}

The proper motions of bright sources ($G\lesssim 13$) in \gdrtwo\ have
a systematic residual reference frame rotation of $\approx 0.15$~\masyr,
whereas faint sources do not show any significant spin relative to quasars
\citep{gaiadr2-astrometry}.
Because the cluster and its tails are spread over a large area on the sky,
this could potentially inject a fake linear velocity field signal. We tested
for any effect this might have on the inference by comparing each fit with
and without the correction for bright sources. We apply the correction for
rotation provided by Lindegren\footnote{Available
on the \gaia\ DR2
\href{https://www.cosmos.esa.int/documents/29201/1770596/Lindegren_GaiaDR2_Astrometry_extended.pdf/1ebddb25-f010-6437-cb14-0e360e2d9f09}{known
issues web page}, slide 32} to sources brighter than $G=12$.
We found that for the Hyades, the systematic \gaia\ frame rotation for bright sources has negligible effect on all parameters.

\subsubsection{Comparison to HARPS study by \citet{leao2019}}

Recently, \citet{leao2019} compared spectroscopic RVs measured from
High-Accuracy Radial velocity Planet Searcher (HARPS) spectra with the
astrometric RVs for 71 stars in the Hyades cluster. They found that the RV
difference is skewed and dependent on the positions of the stars on sky. They
attributed this to cluster rotation of $42.3\pm4.0$~\mspc. We first note that
rotation of such magnitude would easily be revealed by the current method and
data. However, our modelling of \gdrtwo\ astrometry and partial RVs suggests
that there is no significant rotation (\figname~\ref{fig:T_gal}) in the cluster.
The strongest signal we find in the linear velocity gradient is that of
positive shear along the Galactic radial direction. We find that the inferred
shear without any rotation can also produce the $\Delta RV$ versus right
ascensions trend seen by \citet{leao2019}.

\subsubsection{Effects of Binarity and Spurious Astrometry}
\label{sec:binaries}

Binary stars tend to bias the velocity dispersion determined from (single-epoch)
spectroscopic RVs to a larger value as they may include jitters due to binary
orbital motion on top of intrinsic dispersion.
However, it is important to note that the RVs and RV errors reported in
\gdrtwo\ are not from a single-epoch measurement, but the median and scatter
around the mean of multiple per-transit RV measurements for each source
\citep{gaia-dr2-rvvalidation}.
Thus, binarity makes the \gdrtwo\ RV \emph{errors} larger, which will bias the
velocity dispersion to a smaller value (\sectionname~\ref{sub:validation})
in opposite to the usual expectation
\footnote{
  In \gdrtwo, stars with RV errors larger than 20~\kms\ are already filtered
  out and not reported but binaries (and multiple systems) with orbital
  motion inducing smaller RV scatter may still be present
  \citep{gaia-dr2-rvvalidation}.
}.
Larger errors also mean that those stars will not drive the fit as
data are weighted by $(1/\mathrm{error}^2)$.
Of course, if the binary orbital motion introduces a large enough shift in
RV, the star may be excluded from the cluster entirely by the mixture model.

We have tested how much the inferred velocity dispersion is affected by
spurious astrometric measurements including those caused by astrometric binaries
by removing the top $\approx 10$\% outliers in
re-normalized unit weight error (RUWE), i.e., 54 out of 400 sources in the
cl sample with $\mathrm{RUWE}>1.396$.
The RUWE is a goodness-of-fit metric for the single-source astrometric model
re-normalized in order to take out the colour and magnitude dependent
systematics present in \gdrtwo.
Because one main astrophysical cause that makes a source deviate from the
single-source model is binarity, it can be used to pick out
candidate binaries with astrometric wobble \citep{belokurov2020-ruwe-binaries}.
It is also the preferred metric to filter out ill-behaved astrometric
sources \citep{gaiadr2-astrometry}
\footnote{Further information is available on the \gaia\ DR2 \href{https://www.cosmos.esa.int/web/gaia/dr2-known-issues\#AstrometryConsiderations}{known issues web page}.}.
We find that the velocity dispersion remains the same and is not driven by
potential astrometric binaries or spurious astrometric measurements.

Based on these considerations, we conclude that binarity or spurious astrometry are
unlikely to significantly bias the velocity dispersions.
We note that our velocity dispersion estimate of the cluster (the cl sample)
is compatible with the isotropic dispersion determined in previous studies
using different methods and data \citep[][$\sigma_\mathrm{1D}\approx
0.3~\kms$]{reino2018-tgas-hyades,lindegren2000},

\subsubsection{Effects of underestimated errors}

A bug in the astrometric processing software \citep[``DOF bug'',][]{gaiadr2-astrometry}
resulted in serious underestimation of \gdrtwo\ astrometric uncertainties.
While it has been corrected ad hoc at a later stage of the processing,
validation with external data show that they are still underestimated
\citep{gaia-dr2-overview,gaia-dr2-validation}.
The degree of underestimation depends on source magnitudes and varies from
10\% to 50\%.
In an independent investigation, \citet{brandt2018-hipparcos-gaia} found
a similar conclusion: cross-calibrating \gdrtwo\ with \hipparcos\
they find a global multiplicative error inflation factor of $1.743$ for
the proper motions. Moreover, they find that how much the reported errors
are underestimated is spatially varying.

Inference of internal dispersion is degenerate with and dependent upon
correct observational uncertainty estimates, as both work to add noise
to the proper motions except the former is intrinsic to the cluster.
This, combined with binaries and spurious astrometric sources, may
bias the inferred velocity dispersion high.
We tested whether the velocity dispersion changes when we account for both simultaneously
by first removing top 10\% RUWE outliers (\sectionname~\ref{sec:binaries}) and then inflating the errors of
parallaxes and proper motions for all sources by a factor of 2.
Even in this rather extreme scenario of error underestimation, we find no
significant difference in our inference of the internal velocity dispersion.
\gdrtwo\ astrometry for these nearby stars are precise enough
to resolve $\approx 0.4$~\kms\ internal dispersion (see also \figname~\ref{fig:futurecl}).

\section{The Present and Future of the Hyades}
\label{sec:models}

We now discuss our kinematic results with a view to assessing the present
status and future prospects of the Hyades cluster and tails.

\subsection{Steady-state Dynamical Models}

First, let us build a steady-state dynamical model of the Hyades cluster,
inspired by the observation that the light profile in the inner parts follows
a Plummer model \citep{Gu88,roser2011}.
It has long been known that the shape of the Hyades is flattened along the
Galactic $Y$ and $Z$-directions and elongated along the $X$-direction toward
the Galactic Centre
\citep[\figname~\ref{fig:hyades_dist};][]{oort1979,perryman1998,roser2011}.
This is consistent with the effect of the Galactic tides.
Using \citet{reino2018-tgas-hyades}, the Hyades has a prolate shape with axis
ratio $q \approx 0.8$ at the tidal radius of $\rt \approx 10$ pc. This
suggests a model with potential
\begin{equation}
    \phi(x,y,z) = -{GM \over \Bigl((a^2 + r_{\rm c}^2)^2 + 2b^2(y^2+z^2)\Bigr)^{1/4} }.
\end{equation}
Here, ($x,y,z$) are cluster-centric analogues of the Sun-centred $(X,Y,Z)$ coordinates, whereas $r_{\rm c}^2 = x^2 + y^2 + z^2$.
When $b=0$, this is recognised as the familiar \citet{Pl11} sphere.
We use Poisson's equation to obtain the density $\rho(x,y,z)$. For $b \neq
0$, it corresponds to a prolate Plummer spheroid with a long axis in the
$x$-direction and two short axes in $y$ and $z$. Using the result from
\citet{roser2011}, we set the Plummer scale length $a$ as 3.1 pc and choose
the total mass $M$ so that the central density is 2.21 $M_\odot$ pc$^{-3}$.
The axis ratio at the tidal radius is
\begin{equation}
    q = 2^{1/4}(5a)^{5/2}{300a^4 + 164a^2b^2 -9b^4\over (3a^2 +2b^2)(50a^2 +9b^2)^{9/4}}
\end{equation}
So, the scale length $b$ is taken as $2.08$ pc to yield the desired axis
ratio of the density contours as $q \approx 0.8$ at the tidal
radius~\citep{reino2018-tgas-hyades}. This gives us a good representation of
the Hyades stellar density within $\rt = 10$ pc.

It is now simple to solve the (steady-state) Jeans equations which relate the
velocity dispersions to the gravity field of the cluster. As the dispersion
tensor of the cl sample is close to alignment in the ($x,y,z$) coordinates
(see Table~\ref{tab:result} and \figname~\ref{fig:T_gal}), we set the
cross-terms to zero at outset, so the three Jeans equations simplify to
\begin{equation}
    {\partial \rho \sigma^2_{x_i}\over \partial {x_i}} = -\rho {\partial \phi \over \partial x_i}, \qquad\qquad x_i =(x,y,z).
\end{equation}
The position-dependent velocity dispersions are then mass-weighted within the
tidal radius to obtain ($\sigma_x,\,\sigma_y,\,\sigma_z$) =
($0.183,\,0.173,\,0.173)$~\kms. These can be compared with the numbers in
Table~\ref{tab:result}.

The velocity anisotropy of the dynamical model $\sigma_x/\sigma_y$ is 1.06.
From the fitting of the cl sample ($r_c<10$~pc), we infer
$\sigma_x/\sigma_y=1.2^{+0.20}_{-0.22}$.
Thus, while it is of mild significance mainly due to the lack of high-precision
RVs for the bulk of the sample, the slight elongation of the velocity ellipsoid
along $x$-axis is consistent with the dynamical model.
The flattening of the cluster can be explained by the velocity anisotropy observed
and does not require rotation, which is not detected.
However, the total three dimensional velocity dispersion of the dynamical
model is 0.305~\kms (mass-weighted over the cluster within the tidal
radius).
This is a factor of $\approx 2$ smaller than the inferred total three
dimensional velocity dispersion from the kinematical analysis,
$\sqrt{\sigma^2_x + \sigma^2_y + \sigma^2_z} = 0.692$~\kms\
(Table~\ref{tab:result}), leading to a factor of $\approx 4$ discrepancy in
total mass. In fact, using the measured dispersions, our model suggests a 
mass of the Hyades within its tidal radius of $900^{+250}_{-220} \msun$.

It is instructive to compare these results with \citet{roser2011}, especially
their Table 3.
Their three-dimensional velocity dispersion for stars within
the tidal radius is $0.77\pm 0.10$~\kms, in agreement with the
results in Table~\ref{tab:result}.
For a theoretical prediction, they find $0.36$~\kms using the virial
equation, essentially a cruder form of the Jeans equations used here.
So, they also find a discrepancy by a factor of $\approx 2$ between
observations and steady-state models.
They ascribe the disparity mainly to possible inflation of the dispersion
caused by binaries while some hidden mass in white dwarfs, low mass stars,
and binary companions may increase the observed mass by less than 50\%.
However, we argued in \sectionname~\ref{sec:binaries} that the velocity dispersion
we determine is not significantly biased due to binarity.
We conclude that the measured velocity dispersions of the Hyades stars --
even within the tidal radius ($\approx 10$~pc) -- are much
too high. This must be caused by a dynamical mechanism completely absent from
a steady-state Jeans modelling. By extension, {\it the Hyades cluster cannot
possibly be in virial equilibrium and must be close to its disruption}.

If a cluster is in a perfectly circular orbit in an axisymmetric potential,
the gravitational potential in the frame co-rotating with the cluster is
static and tidal heading would be irrelevant. Such conditions are never met
in reality. The epicyclic motion of the cluster introduces
time-variation, which results in tidal heating.
While repeated tidal shocks due to encounters with molecular clouds may also
heat the cluster, the preferential direction in which the velocity dispersion
is larger in both cl an tails fits naturally seem to prefer the Galactic
tides as the explanation.
Once stars escape from the cluster beyond the tidal radius of $\approx 10$~pc
for the Hyades, the escaped stars follow their own epicyclic motions.
This, combined with the fact that tidal heating is even more effective for
the stars in the tails that are free from the cluster potential, may explain
the even larger velocity dispersion along the $x$-direction for the tails
fits.

\subsection{Evolving Models}
\label{sec:evolving}

\begin{figure*}
  \centering
  \includegraphics[width=0.85\linewidth]{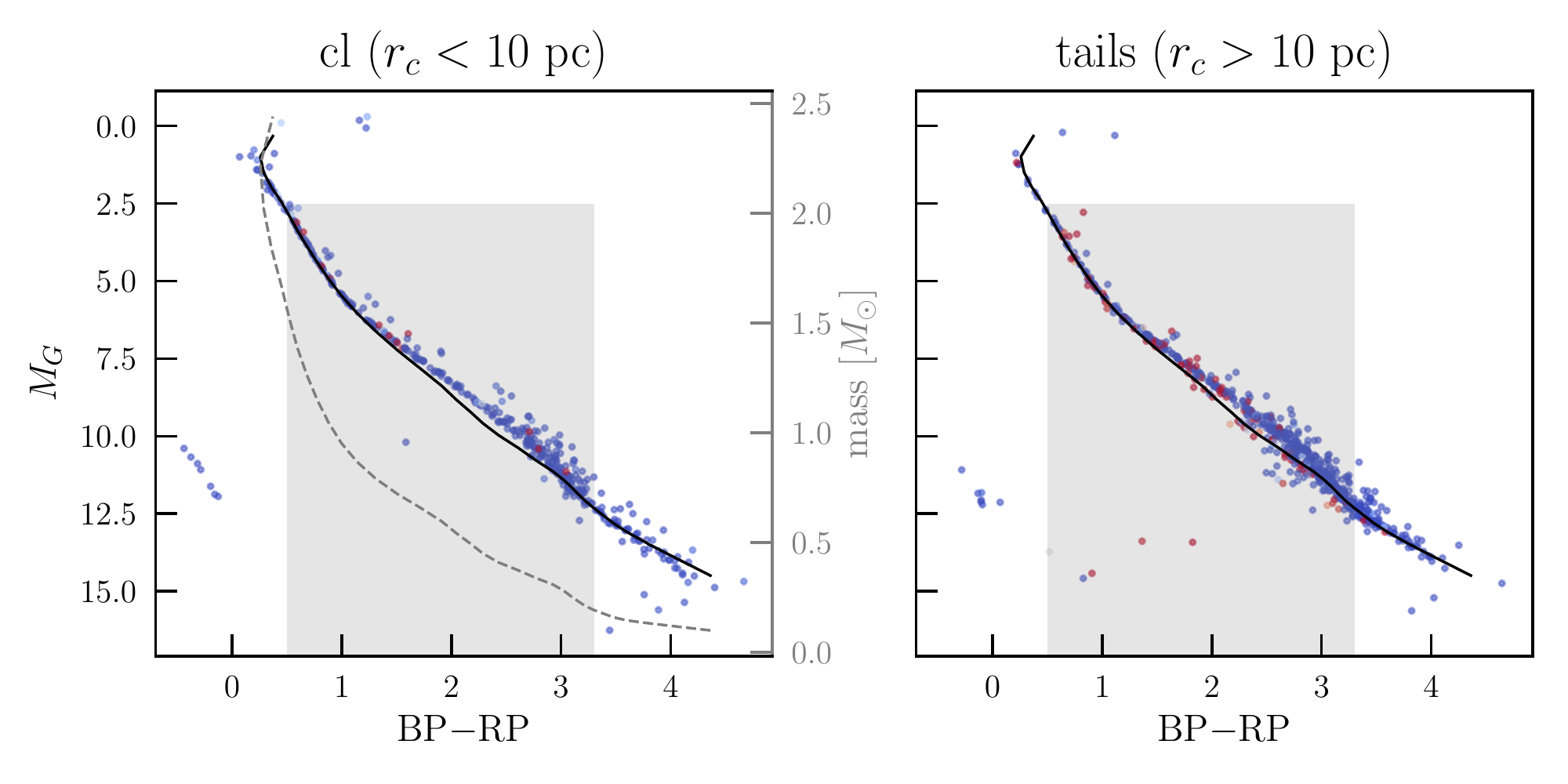} 
  \caption{
    Color-magnitude diagrams of the cluster and tails.
    Sources are colored by their mean membership probability
    as in \figname~\ref{fig:membership}.
    In order to estimate \etanow, we convert the \bprp{} color
    to mass using the MIST isochrone
    \citep{choi2016-mist} of 680~Myr and $\feh=0.24$ \citep{gossage2018}.
    The isochrone in the color-magnitude space is the black line in each
    panel.
    The color-mass relation is shown in the left panel as the gray dashed
    line with the corresponding y-axis on the right.
    Only sources within the shaded region
    ($0.5<\bprp<3.3$ corresponding to $0.19<m<1.56$~$M_\odot$)
    are considered (see \sectionname~\ref{sec:evolving}).
  }
  \label{fig:cmd}
\end{figure*}

To build evolving models of the Hyades, we assume that the tidally-stripped stars leave with nearly zero energy.
Then, the mass $M(t)$ and Plummer scale length $a(t)$ of the cluster decline with time, while the cluster energy $E$
\begin{equation}
E = -{3\pi\over 64} {GM^2(t)\over a(t)}.
\end{equation}
remains constant. As \citet{Ku08} point out, this is a surprisingly good approximation as stars cross the tidal radius with almost zero velocity. 

We assume that mass is lost according to \citep[cf][]{He61,Gn99}
\begin{equation}
{dM\over dt} = -{k M\over t_{\rm r}}
\label{eq:ode}
\end{equation}
where $k$ is an unknown constant and $t_{\rm r}$ is the relaxation time at the half-mass radius \citep[see e.g.,][]{SpH}, which for the Plummer model is
\begin{equation}
t_{\rm r} = 0.206 {M(t)^{1/2} a(t)^{3/2} \over G^{1/2} \langle m_\star \rangle \log \Lambda}.
\end{equation}
Here, $\langle m_\star \rangle$ is the mean stellar mass, while $\Lambda$ is the Coulomb logarithm.
This ansatz~(\ref{eq:ode}) encodes the complicated physics of evaporation and ejection of stars, tidal stripping and disc shocking.
Although simple, it has been used with success to represent the results of full N-body simulations of globular clusters, with values of $k$ in the range 0.05 - 0.007~\citep[e.g.,][]{Sp73,Gn99}.

This differential equation (\ref{eq:ode}) can be solved~\citep[][chap 7]{BT} to give power-law solutions for the mass of the cluster $M(t)$, its scale length $a(t)$ and its tidal radius $r_{\rm t}(t)$.
Specifically, we obtain:
\begin{equation}
M(t) = M_0 \left( 1 - {7 k t\over 2 t_{r,0}}\right)^{2/7},\qquad
a(t) = a_0 \left( 1 - {7 k t\over 2 t_{r,0}}\right)^{4/7},
\label{eq:timecurves}
\end{equation}
where $t$ is the time since formation, whilst a zero subscript indicates the value at the initial time.

The age of the Hyades cluster is $\tnow \approx 680$~Myr \citep{gossage2018}.
Its present day stellar mass is $M(\tnow) = \Mnow \approx 300$ $M_\odot$, whilst its relaxation time $\trnow$ is $\approx 60$~Myr.
By themselves, these data are not enough to prescribe the location of the Hyades on the evolutionary tracks given by eqs~(\ref{eq:timecurves}).
However, there is a further piece of information that is susceptible to observational scrutiny, namely the ratio of the stellar mass in the tails to the mass in the cluster
\begin{equation}
    \eta(t) = {M_0 - M(t) \over M(t)}. 
\end{equation}
Although this is not precisely known, it is evident that the tails are mature and well-developed (see \figname~\ref{fig:hyades_dist}).

In order to estimate \etanow\ from the data,
we use the MIST model isochrone \citep{choi2016-mist} of
680~Myr and $\feh=0.24$ \citep{gossage2018} to convert \bprp\ color
to mass.
\figname~\ref{fig:cmd} shows the distribution of sources for the cluster and tails
along with the model isochrone.
The color-mass conversion curve is also shown in the gray dashed line in the left panel.
We only consider stars within the color range $0.5<\bprp<3.3$ indicated
as the shaded gray region in \figname~\ref{fig:cmd}.
This corresponds to $0.19<m<1.56$~$M_\odot$ in mass.
The blue limit is set so as not to deal with the main-sequence turn off,
which makes color-mass relation non-monotonic.
At the red color limit, the observed magnitude is $G\approx 15-16$, well below
the \gaia\ magnitude limit ($G\approx 20$).
When summing up the mass of stars, we only include stars with mean membership
probability larger than $0.5$ taking advantage of the kinematic membership
we infer in our kinematic modelling.
We find that the cluster and tails contain $157$~\msun\ (288 sources)
and $239$~\msun\ (587 sources) respectively at this color (mass) range,
resulting in our crude \etanow\ estimate as $\etanow \gtrsim 1.5$.
This is likely a lower bound as the extent of tails discovered and included
in our study is not complete.
Note also that the model isochrone is not well matched to the data at low
mass and that we have not taken binarity of sources into account in
converting color to mass.
A more careful modelling is required to refine individual sources' mass estimate.

In any case, the Hyades has already undergone substantial destruction as
\etanow\ is evidently larger than 1.
The original mass of the Hyades cluster at birth is
\begin{equation}
   M_0 = \Mnow ( 1 + \etanow) \gtrsim 750~\msun.
\end{equation}
By using $\eta$ as a proxy for time, we find the present day mass loss rate of stars is
\begin{align}
    -{d M\over dt} &= {2 ((1+\etanow)^{7/2} -1)\over 7} {\Mnow \trnow \over \tnow^2} \\
    &\gtrsim 0.26~\msun\,{\rm Myr}^{-1}.\nonumber
\end{align}
These results can be compared with the numerical simulations of
\citet{ernst2011}, who attempted to reproduce the present-day cumulative mass
profile, stellar mass and luminosity function of the Hyades. Their
best-fitting Plummer model has an initial mass of $1230\, M_\odot$ and an
average mass loss rate of $1.4\, M_\odot$Myr$^{-1}$.

The mass loss rate will increase rapidly as the cluster approaches complete disintegration. The lifetime of the Hyades is
\begin{equation}
t_{\rm life} = \tnow \left({1\over 1 - (1+\etanow)^{-7/2}}\right) \lesssim 709~{\rm Myr} 
\label{eq:lifet}
\end{equation}
As the age of the Hyades is $\approx 680$ Myr, the cluster is in its death throes.
The final dissolution will take place over the next $\lesssim 30$ Myr.
The end is nigh for the Hyades irrespective of the precise value of
$\etanow$, providing it is $\gtrsim 1$.
This is because all such curves (\ref{eq:lifet}) decline precipitously at
late times and the bound mass drops quickly to zero.
At dissolution, the stars of the Hyades are all unbound, but they have not
necessarily dispersed from the location of the object.
Of course, as the tails continue to stretch and evolve, the stars disperse
and the remnant itself becomes indistinguishable from the tails (cf the
dissolution of Ursa Major II discussed in \citet{Fe07}, especially their
Figure 6).

Note that both our steady-state and evolving models tell a consistent story.
The Hyades cluster is far from any virial equilibrium, and it cannot be
expected to survive in it present fragile state for much longer. This result
is implicit in earlier work~\citep{roser2011,ernst2011}, but the phenomenal
quality of the \gaia\ proper motions allow us to be much more explicit here.

\subsection{The Hyades Tails}

The planar velocity field in the tails relative to the systemic motion of the
cluster can be described in terms of analogues of the Oort Constants
as~\citep{Oo28,Og32,Mi35, Mu83}
\begin{equation} \label{eq:oort-def-b}
  \mathbf{T_2} = \left(
    \begin{array}{c@{\;\;}c}
      \partial v_x/\partial x & \partial v_y/\partial x \\ 
      \partial v_x/\partial y & \partial v_y/\partial y
    \end{array} 
  \right) \equiv \left(
    \begin{array}{c@{\;\;}c}
      K+C & A-B \\
      A+B & K-C
    \end{array}
  \right).
\end{equation}
The parameters $A$, $B$, $C$, and $K$ are the Oort Constants and they measure
the local (two-dimensional) divergence ($K$), vorticity ($B$), azimuthal
($A$) and radial ($C$) shear of the velocity field. By comparison with
Table~\ref{tab:result}, we see that
\begin{eqnarray}
\label{eq:oorttails}
    A &= w_3 = 16.90 \pm 0.92\, \mspc,\nonumber\\
    B &= \omega_z = -6.48 \pm 1.15\, \mspc,
\end{eqnarray}
\begin{eqnarray}
    C &= \frac{1}{2}(w_4 - w_5) = 1.78 \pm 1.27\,\mspc \nonumber\\
    K &=\frac{1}{2}(w_4 + w_5) = 2.49\pm 1.27\,\mspc
\end{eqnarray}
These may be compared to the Oort constants describing deviations of the
velocity field from the Local Standard of Rest. Using the \tgas\ catalogue,
\citet{Bovy} found: $A = 15.3 \pm 0.4$ \mspc, $B = -11.9 \pm 0.4$ \mspc, $C =
-3.2 \pm 0.4$ \mspc and $K = -3.3 \pm 0.6$ \mspc. Exact agreement is not
expected, as (i) the Taylor expansion is around the mildly eccentric Hyades
orbit rather than a cold circular orbit in the Galactic plane ($Z=0$), (ii)
processes like the mass loss, disk and bulge shocking and perturbations by
molecular clouds and spiral arms may also affect the kinematics of the tidal
tails.

Using Table~\ref{tab:result}, the velocity dispersion of tail material has
$\sigma_y / \sigma_x = 0.64 \pm 0.06$ and $\sigma_z /\sigma_x = 0.48 \pm
0.04$. These results may be compared to the analogous values for the thin
disk ($\sigma_y / \sigma_x = 0.70 \pm 0.13$ and $\sigma_z /\sigma_x = 0.64
\pm 0.008$) and the thick disc ($\sigma_y / \sigma_x = 0.67 \pm 0.11$ and
$\sigma_z /\sigma_x = 0.66 \pm 0.11$) found with \gaia\ DR1 by \citet{Ang18}.
Note that steady-state populations in an axisymmetric disc with a flat
rotation curve are predicted to have $\sigma_y / \sigma_x = \sqrt{2} = 0.707$
in epicyclic theory~\citep[e.g.,][]{Ev93,KT}, very close to what is seen for
the thin disc. The Hyades tail stars possess kinematics distinct from both
discs with the in-plane dispersion ratio comparable to the thick disc, but
the vertical ratio much colder. The vertex deviation $\ell_{uv}$ and tilt
angle $\ell_{uw}$ are defined as~\citep[e.g.,][]{Smith}
\begin{equation}
    \ell_{uv} = {1\over2}\arctan \left ( {2\Omega_{xy} \over \sigma^2_x-\sigma^2_y}\right),\quad
   \ell_{uw} = {1\over2}\arctan \left ( {2\Omega_{xz} \over \sigma^2_x-\sigma^2_z}\right),
\end{equation}
Both angles should vanish for a relaxed stellar population in an axisymmetric disc~\citep[e.g.,][]{SEA}.
However, the values of the cross terms in Table~\ref{tab:result} betray significant
vertex deviation and tilt for the Hyades tail population, and we calculate
$\ell_{uv} = 34^\circ$ and $\ell_{uw} = 19^\circ$. These are very different
from the thick disc, which has a roughly constant vertex deviation of
$\ell_{uv} \approx -15^\circ$ and tilt $\ell_{uw} \approx -10^\circ$. They
are however similar to the vertex and tilt of the thin disc stars with
comparable metallicity ([Fe/H] $\approx 0.24$), as shown in Figures 11 and 12
of \citet{Ang18}. Nonetheless, the axis ratios and misalignment angles
together demonstrate that the kinematic properties of the tail material do
differ from those of the field population, which may enable efficient
filtering of tail stars to aid detection of material at larger distances from
the cluster.

\section{Summary}

\begin{figure}
  \centering
  \includegraphics[width=0.95\linewidth]{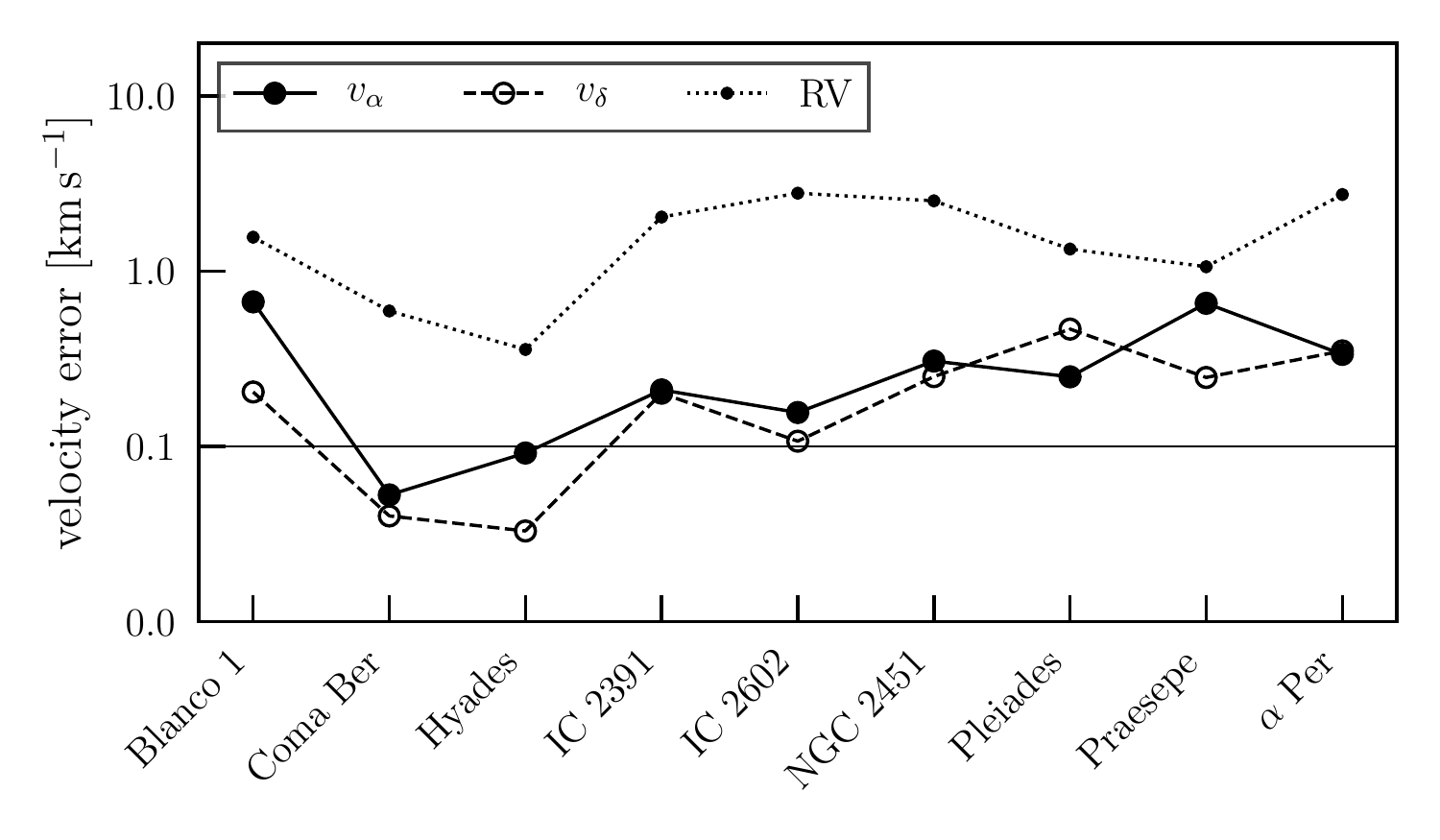} 
  \caption{
    Median \gdrtwo\ velocity errors for a number of nearby open clusters and
    associations \citep{gaiacollab-dr2-hrd}.
    }
  \label{fig:futurecl}
\end{figure}

The unprecedented quality of the \gaia\ data and its synergy with various
spectroscopic surveys have already started to improve dramatically our
understanding of star clusters. Internal kinematics of clusters in particular
can provide valuable hints as to their formation, evolution and destruction.

We presented a method to model the internal kinematics of stars in a cluster
or association, which builds upon and extends previous works with \hipparcos\
and \gaia. Our model allows for anisotropic velocity
dispersions and a linear velocity gradient (equivalently, rotation and
shear).
It incorporates radial velocity measurements for a subset of stars with
astrometry, and accounts for contamination by background sources (in terms of
velocity) via a mixture model.
We implemented the method in a modern statistical modelling language and
validated the implementation with mock data generated with similar quality as
\gdrtwo.

We applied the method to the \gdrtwo\ data of the Hyades cluster and its
tails, which have recently been discovered using the same
data~\citep{meingast2019,roser2019}. We divided the sample into two, the
cluster proper (cl, $r_c<10$~\pc) and the tidal tails (tails, $r_c>10$~\pc).

While the velocity dispersion of the cluster is nearly isotropic,
there is slight elongation of the velocity ellipsoid in the Galactic radial
direction consistent with what is expected from the prolate shape of the cluster.
The Hyades shows no evidence for internal rotation.
Strictly speaking, this result is restricted to solid body rotation and other
rotation laws are possible for clusters~\citep[e.g.,][]{Er07,Je17}, though in
the cluster centre they do reduce to solid body rotation. There is almost no
net expansion or contraction of the cluster stars, but there are measurable
($\approx 2\sigma$) negative and positive gradients in $dv_y/dy$ and
$dv_z/dz$ respectively. The shear without any rotation can produce the trends
seen by \citet{leao2019} and claimed as rotation.

The stars in the tidal tails ($r_c>10$~pc) show clear
velocity dispersion anisotropy and linear velocity gradient.
The kinematics of the tail stars parallel to the Galactic plane can be
decomposed into a shear of $16.90\pm 0.92$ \mspc and a vorticity of $-6.48\pm
1.15$ \mspc. These values are different from the local Oort constants of $A =
15.3 \pm 0.4$ \mspc and $B = -11.9 \pm 0.4$ \mspc~\citep{Bovy}. This is
because the velocity gradients are measured with respect to the Hyades
cluster orbit, which is mildly elliptical with an eccentricity of $\approx
0.1$ and a vertical libration amplitude of $\approx \pm 70$ pc
~\citep{ernst2011}. The classical Oort constants apply in the cold limit of
vanishing random motions, in which the mean streaming velocity is the
velocity of closed circular orbits supported by the Galactic potential.

The Hyades cluster has a prolate shape, fashioned by the Galactic tides. It
is flattened along the $y$ and $z$-directions, but elongated along the
$x$-direction towards the Galactic centre. The stellar density of the cluster
is well modelled by a prolate Plummer spheroid. By solving the Jeans
equations, we find the velocity dispersions needed to support the cluster in
virial equilibrium. These are a factor of $\approx 2$ smaller than the values
inferred from our kinematic analysis. It follows that the Hyades cluster is
not in virial equilibrium. Many of the stars in our cl sample are unbound.
Their velocities are strongly enhanced by tidal heating, providing the
population of ``potential escapers'' identified by \citet{Ba01} and \citet{kupper2010}. 

Simple models of Hyades evolution driven by mass loss are then developed.
Assuming the stripped stars leave with almost zero energy with respect to the
cluster, the total mass and tidal radius behave like power-laws of the
time until dissolution. Given an estimate of the ratio of the mass in the
tails to the mass in the cluster at the present epoch, then the original
mass, the current mass loss and the time till disruption can be computed.
Using the extent of tidal tails so far discovered,
we can place a lower bound on the initial mass of the Hyades at birth
and its current mass loss rate as
$M_0 \gtrsim 750~\msun$ and $-dM/dt \gtrsim 0.26 \msun\,\mathrm{Myr}^{-1}$.
We estimate that less than $30$~Myr is left until the final dissolution;
the Hyades is in its death throes. 

There are a number of avenues for future exploration. Although our kinematic
model has a global velocity dispersion matrix for the cluster, in practice it
changes as a function of distance from the cluster centre. For example, in
Plummer models of the Hyades, the velocity dispersions fall by a factor of
1/3 on moving from the centre to the tidal radius. Thus, a better kinematic
model might be one in which the dispersion matrix changes as a function of
cluster-centric distance. As this introduces more free parameters, it
inevitably increases the uncertainties on inferred parameters.

Secondly, much remains to be done to complete the census of Hyades tail
members as the searches so far have been limited to 200~pc from the Sun
\citep{roser2019,meingast2019}.
We can use the shear to estimate the total length of the tidal tails
of the Hyades. Given its age of $\tnow \approx 680$ Myr, then the first stars
to stripped will now lag or lead the cluster by a distance
\begin{equation}
    L\approx 2A\, {\sqrt{3}\rt}\,\tnow \approx 400\, {\rm pc},
\end{equation}
where we have taken the Oort constant $A$ for the tails from
eq~(\ref{eq:oorttails}) and assumed the tails are displaced in radius from
the cluster by $\sqrt{3}\rt$ using eq.~(17) of \citet{Ju09}.
The simulations of \citet{ernst2011} found a somewhat larger value of
$L\approx 800$ pc.
Of course, it becomes increasingly challenging to trace convincingly
the low-density tidal tails as the distance from the cluster centre increases.

The most favourable locations at which to look for extensions of the Hyades
tails are the overdensities caused by epicyclic bunching of tail
stars~\citep{Ku08, Ju09}. Once stars leave the Hyades, their motion is
well-described by epicyclic theory. The relative velocity of tail stars is
then smallest at the pericentres of the epicycles. The locations of epicyclic
clumpings are~\citep{Ju09}
\begin{equation}
y \approx  \pm {4\pi \Omega\over \kappa}{4\Omega^2 -\kappa^2\over \kappa^2}\sqrt{3}\rt
\end{equation}
where $\Omega$ and $\kappa$ are the circular and epicyclic frequencies. For
the Hyades, this gives $y \approx \pm 154$ pc, taking $\kappa/\Omega =
\sqrt{2}$ appropriate for a flat rotation curve. This clumping phenomenon has
only been seen in simulations~\citep{Ku08,Ju09}, but has not been
unambiguously shown to occur in nature. In fact, perturbations from spiral arms
or giant molecular clouds may complicate the picture from epicyclic theory
and disperse such density enhancements, rendering them undetectable in
practice.

Finally, although we have concentrated on the Hyades cluster in this paper,
it is natural to extend the work to other nearby open clusters with high
quality data. \figname~\ref{fig:futurecl} shows median velocity errors
derived from \gdrtwo\ spectroscopic and astrometric
measurements for a sample of nearby open clusters \citep{gaiacollab-dr2-hrd}.
In terms of precision of the velocities, the Hyades is the most
propitious target, with Coma Berenices, IC~2602 and Praesepe being the next
most favourable. The number of members in \gdrtwo\ varies from $\approx 150$
for Coma Berenices~\citep{Ta18} to $\approx$ 1500 for the
Pleiades~\citep{Ga19}. Some of these open clusters also have newly identified
tails~\citep{RoSc19, Ta19}. This opens up the possibility of using the tails
of nearby open clusters to measure kinematic properties at multiple locations
in the Galaxy.

The data underlying this article are available in the article and in its
online supplementary material.
The modelling code is available at \url{http://github.com/smoh/kinesis}.

\section*{Acknowledgments}

Comments from the Cambridge Streams Group are gratefully acknowledged.
We thank the anonymous reviewer whose comments helped improve and
clarify this manuscript.
This work has made use of data from the European Space Agency (ESA) mission
{\it Gaia} (\url{https://www.cosmos.esa.int/gaia}), processed by the {\it Gaia}
Data Processing and Analysis Consortium (DPAC,
\url{https://www.cosmos.esa.int/web/gaia/dpac/consortium}). Funding for the DPAC
has been provided by national institutions, in particular the institutions
participating in the {\it Gaia} Multilateral Agreement.
This project was developed in part at the 2019 Santa Barbara Gaia Sprint,
hosted by the Kavli Institute for Theoretical Physics at the University of
California, Santa Barbara. This research was supported in part at KITP by the
Heising-Simons Foundation and the National Science Foundation under Grant No.
NSF PHY-1748958.
This work made use of matplotlib \citep{matplotlib:2007},
numpy \citep{numpy:2011}, 
scipy \citep{scipy2019}, astropy \citep{astropy2013,astropy2018},
pandas \citep{pandas:2010}, stan \citep{carpenter2017} and pystan.




\bibliographystyle{mnras}
\bibliography{kinesis} 



\appendix

\section{Perspective effect}
\label{appendix:perspective-effect}

\begin{figure*}
  \centering
  \includegraphics[width=0.7\linewidth]{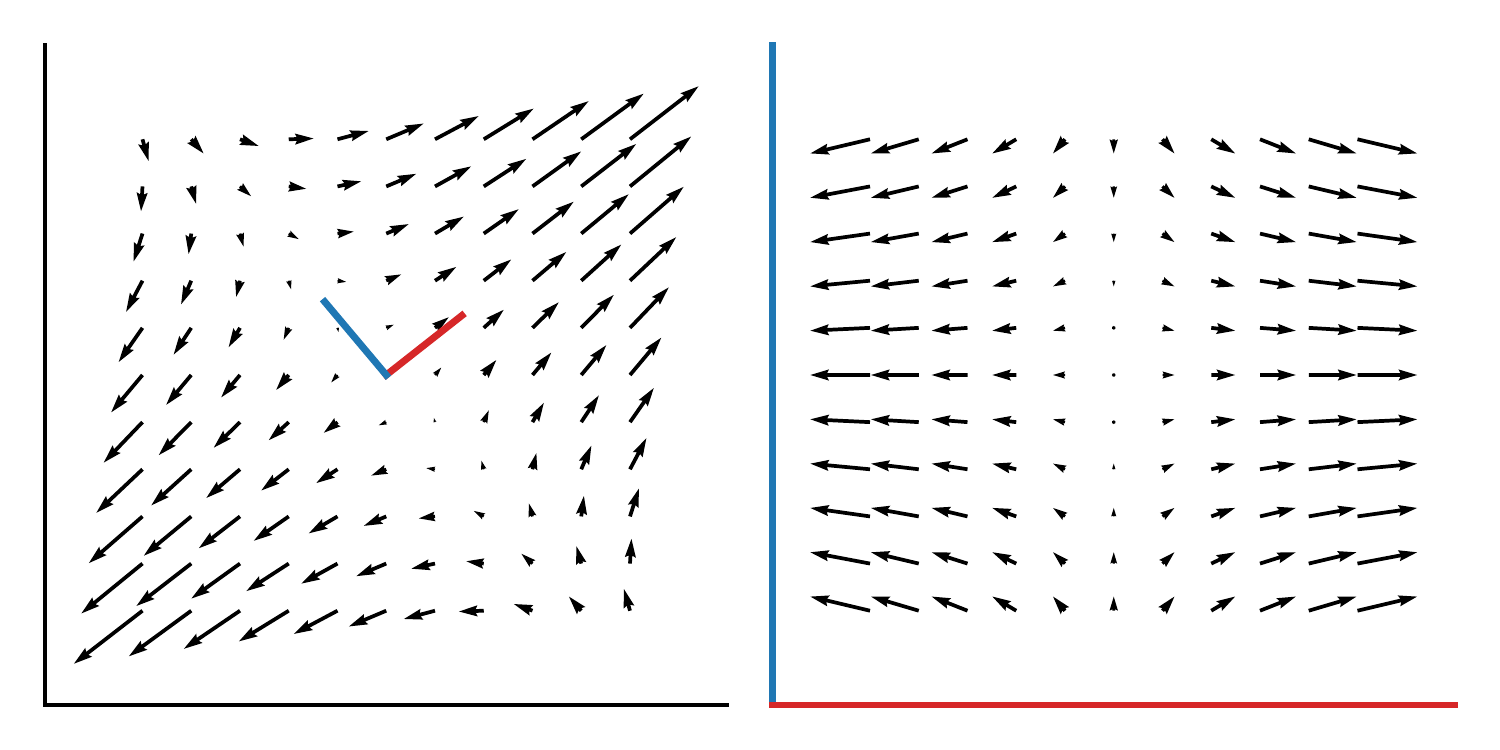} 
  \caption{
    Shear is expansion rotated. The shear is the symmetric part of the
    velocity gradient matrix and so can be diagonalised in a suitably chosen frame
    in which the motion appears as expansion or contraction along the axes.
  }
  \label{fig:rotating-shear}
\end{figure*}

\begin{figure*}
  \centering
  \includegraphics[width=0.7\linewidth]{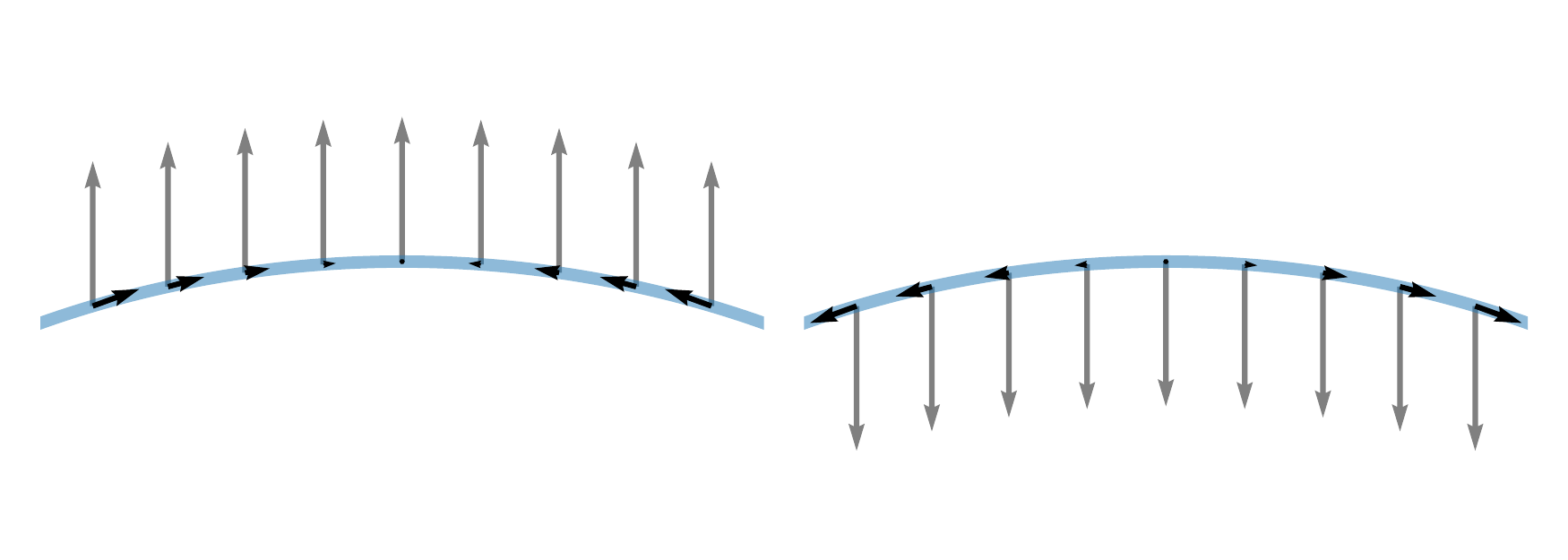} 
  \caption{Receding clusters appear to be contracting while
    approaching clusters appear to be expanding.
    The blue circle is the celestial sphere.
  }
  \label{fig:perspective-radial}
\end{figure*}

\begin{figure*}
  \centering
  \includegraphics[width=0.95\linewidth]{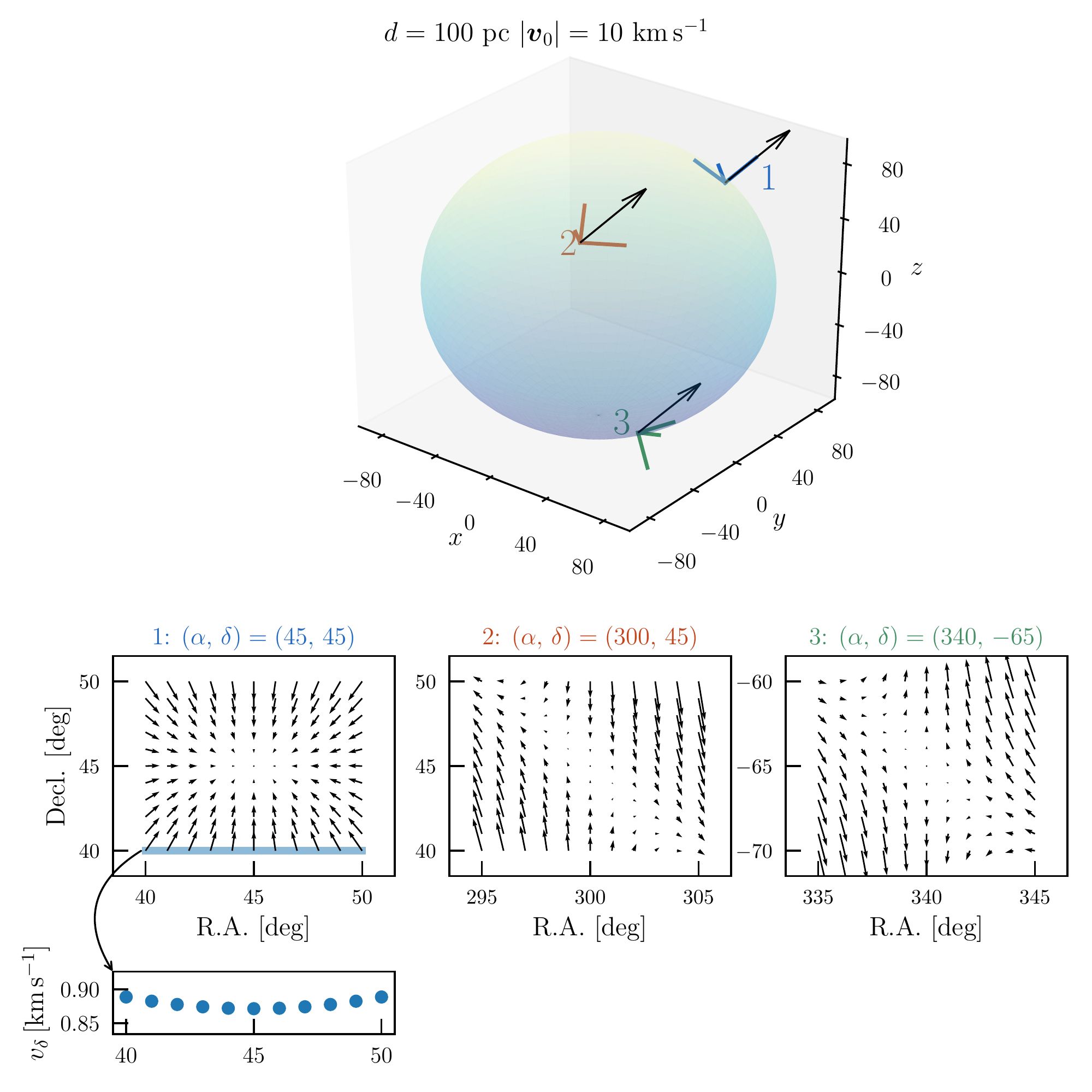} 
  \caption{
    Perspective effect of the same mean velocity vector $\vec{v}_0^T =
    (v_x,\,v_y,\,v_z)=(5.0, 5.0, 7.071)~\kms$ at three different sky
    positions.
    The top 3D plot shows the mean velocity vector at the three different positions as black arrows.
    At each position, the cluster is assumed to be at $d=100$~pc, and the
    basis vectors for the tangent spaces, $\{\vec{p},\,\vec{q},\,\vec{r}\}$,
    are indicated.
    The mean velocity vector was chosen to be radial at position 1, which is
    why it coincides with the basis vector in the radial direction.
    The spherical surface of radius $100$~pc make it clear that
    the basis vectors $\vec{p}$ and $\vec{q}$ span the tangent plane to the sphere
    and $\vec{r}$ is the radial direction normal to that plane.
    Proper motions of stars are due to velocities of stars projected onto the tangent plane.
    Perspective effect arise because of the residuals in the projected space
    due to the curvature of the celestial sphere.
    We show these \emph{residual} projected velocities at each location in the bottom
    sampling a $5^\circ \times 5^\circ$ grid of $(\alpha,\,\delta)$.
    Note that for the same $\vec{v}_0$, the pattern is different at different positions on the sky.
    Generally, it also depends on the values of $\vec{v}_0$ as well.}
  \label{fig:perspective-effect}
\end{figure*}

This appendix aims to illustrate the perspective effect of the \emph{mean}
velocity, $\vec{v}_0$, that may introduce an \emph{apparent} velocity gradient
in the \emph{projected} (observable) position-velocity space.
Note that linear velocity gradients (apparent or real) describe
shear and (solid-body) rotation (see \sectionname~\ref{sub:model}),
where we use `shear' to refer to the general symmetric component of the gradient matrix
$T_{ij} = d v_i / d x_j$ including isotropic contraction or expansion.
Any shear component of $T_{ij}$ can be diagonalized as it is symmetric,
meaning that there is a frame defined by the three principal axes
in which the velocity pattern is either accelerating or decelerating along each axis.
This is illustrated pictorially in \figname~\ref{fig:rotating-shear}.

Because we measure motion of stars on the celestial sphere,
a velocity vector projects differently depending on the star's position.
Thus, in order to model the intrinsic velocity pattern,
we want to set up a coordinate system independent of a star's position on the celestial sphere,
i.e., R.A. and Decl., $(\alpha,\,\delta)$.
The choice of this coordinate system is entirely arbitrary but
given the two angles of spherical coordinates, R.A. and Decl.\footnote{
  Since the declination $\delta$ in astronomy is defined as the angle from the equator,
  it is related to the usual polar angle $\phi$ of spherical coordinate systems
  (angle from $z$-axis) as $\phi = \pi/2 - \delta$.
  We could have easily chosen a different coordinate system to our liking:
  for example, there is another coordinate system defined by two spherical angle pairs,
  namely Galactic longitude and latitude ($l$ and $b$),
  the Galactic coordinates. The choice is entirely arbitrary as long as
  the coordinate transformation is correctly accounted.
}, a natural choice would be the cartesian ICRS defined by these two angles
which we have adopted in this work.

Let us call the velocity in this fixed rectangular coordinate system $\vec{v}_0$
and use $\vec{v}_\mathrm{sphere}$ to refer to the velocity
in the rotated frame defined by two tangent directions along
increasing R.A. and Decl. and one radial direction 
\emph{at $(\alpha,\,\delta)$}, i.e., by the orthonormal basis $\{\vec{p},\vec{q},\vec{r}\}$
we introduced in \sectionname~\ref{sub:model}\footnote{
  Note that a vector is defined by its magnitude and direction, which is independent
  of any coordinate system employed. Here, $\vec{v}_0$ and $\vec{v}_\mathrm{sphere}$ are
  two different representations of the same velocity vector in two different basis sets:
  one that is independent of the position on the celestial sphere and the other that is.
}:
\begin{equation}
  \vec{v}_0 =
    \begin{bmatrix}
      v_x \\
      v_y \\
      v_z
    \end{bmatrix};\qquad
  \vec{v}_\mathrm{sphere} =
    \begin{bmatrix}
      v_\alpha \\
      v_\delta \\
      v_r
    \end{bmatrix}.
\end{equation}
Here, $v_\alpha$ and $v_\delta$ are the velocities along R.A. and Decl. directions
and $v_r$ is the radial velocity.
The first two components are related to the proper motion and parallaxes as
\begin{equation}
  \begin{split}
    v_\alpha &= \mu_\alpha / \pi \\
    v_\delta &= \mu_\delta / \pi.
  \end{split}
\end{equation}
Then, as we state in \sectionname~\ref{sub:model}, the two coordinates of the same velocity vector
are related by a rotational transformation:
\begin{equation}
  v_\mathrm{sphere} = \mat{R}(\alpha,\,\delta) v_0
\end{equation}
where we emphasize that $\mat{R}$ depends on the position $(\alpha,\,\delta)$ on the
celestial sphere.
Of course, $\mat{R}$ is simply
\begin{equation}
  \mat{R} =
  \begin{bmatrix}
   \vec{p}^T \\
   \vec{q}^T \\
   \vec{r}^T
  \end{bmatrix}
  =
  \left[\begin{matrix}- \sin\alpha  & \cos\alpha  & 0\\
    - \sin\delta  \cos\alpha  & - \sin\alpha  \sin\delta  & \cos\delta \\
    \cos\alpha  \cos\delta  & \sin\alpha  \cos\delta  & \sin\delta \end{matrix}\right]
\end{equation}
where we have now given explicit expressions for $\{\vec{p},\vec{q},\vec{r}\}$,
which may be easily obtained by differentiating the usual
rectangular-to-spherical coordinate transformation.
Note that $\mat{R}$ is orthogonal, i.e., $\mat{R}^{-1} = \mat{R}^T$.

The perspective effect arises because of the changing perspective $\Delta(\alpha,\,\delta)$, and
not an actual change in the velocities which we try to infer.
In order to see the lowest order changes, we can expand $\mat{R}$
around a position $(\alpha,\,\delta)$ for some change $(\Delta \alpha,\, \Delta\delta)$,
\begin{equation}
  \begin{split}
    \Delta{\vec{v}_\mathrm{sphere}} &= \Delta{\mat{R} }v_0\\
      &= \left[ \frac{d \mat{R}}{d \alpha} \Delta \alpha + \frac{d \mat{R}}{d \delta} \Delta \delta \right]
        _{(\alpha,\,\delta)} v_0
      + \mathcal{O} (|\Delta \alpha|^2,\, |\Delta \delta|^2,\,|\Delta \alpha \Delta \delta|)
  \end{split}
\end{equation}
Since
\begin{equation}
  \begin{split}
  \frac{d \mat{R}}{d \alpha} &= \left[\begin{matrix}- \cos\alpha  & - \sin\alpha  & 0\\\sin\alpha  \sin\delta  & - \sin\delta  \cos\alpha  & 0\\- \sin\alpha  \cos\delta  & \cos\alpha  \cos\delta  & 0\end{matrix}\right] \\
  \frac{d \mat{R}}{d \delta} &=  \left[\begin{matrix}0 & 0 & 0\\- \cos\alpha  \cos\delta  & - \sin\alpha  \cos\delta  & - \sin\delta \\- \sin\delta  \cos\alpha  & - \sin\alpha  \sin\delta  & \cos\delta \end{matrix}\right],
  \end{split}
\end{equation}
the general expression for how projected velocities change with sky positions is
\begin{equation}
  \begin{split}
  \Delta \vec{v}_\mathrm{sphere} &=
  \begin{bmatrix}
    \Delta v_\alpha \\
    \Delta v_\delta \\
    \Delta v_r
  \end{bmatrix} \\
    &= \left[\begin{matrix}\Delta\alpha \left(- v_{x} \cos\alpha  - v_{y} \sin\alpha \right)\\\Delta\alpha \left(v_{x} \sin\alpha  \sin\delta  - v_{y} \sin\delta  \cos\alpha \right)\\\Delta\alpha \left(- v_{x} \sin\alpha  \cos\delta  + v_{y} \cos\alpha  \cos\delta \right)\end{matrix}\right]\\
    &+ \left[\begin{matrix}0\\\Delta\delta \left(- v_{x} \cos\alpha  \cos\delta  - v_{y} \sin\alpha  \cos\delta  - v_{z} \sin\delta \right)\\\Delta\delta \left(- v_{x} \sin\delta  \cos\alpha  - v_{y} \sin\alpha  \sin\delta  + v_{z} \cos\delta \right)\end{matrix}\right] \\
    &+ \mathcal{O} (|\Delta \alpha|^2,\, |\Delta \delta|^2,\,|\Delta \alpha \Delta \delta|).
\end{split}
\end{equation}
The formula is given not because it is particularly useful but to
show explicitly that generally, the slope of the change in projected velocities
as a function of changing perspective ($\Delta\alpha$ and $\Delta\delta$)
not only depends on sky positions but also on the velocity itself:
perspective effect is why we can infer all three components of the mean velocity of a cluster
by geometry from parallaxes and proper motions (astrometric radial velocity).
We may express this in terms of $\vec{v}_\mathrm{sphere}$ at $(\alpha,\,\delta)$ which
is quite simpler and has been used in practice:
\begin{equation}
  \begin{split}
  \Delta{\vec{v}_\mathrm{sphere}} = &
  \left[ \frac{d \mat{R}}{d \alpha} \Delta \alpha + \frac{d \mat{R}}{d \delta} \Delta \delta \right]
        _{(\alpha,\,\delta)} \left(\mat{R}^T \left[ 
          \begin{matrix} v_\alpha \\ v_\delta \\ v_r \end{matrix}
        \right]  \right)\\
        & + \mathcal{O} (|\Delta \alpha|^2,\, |\Delta \delta|^2,\,|\Delta \alpha \Delta \delta|)\\
        =&
        \left[\begin{matrix}\Delta\alpha \left(v_\delta \sin\delta  - v_{r} \cos\delta \right)\\- \Delta\alpha v_\alpha \sin\delta  - \Delta\delta v_{r}\\\Delta\alpha v_\alpha \cos\delta  + \Delta\delta v_\delta\end{matrix}\right].
  \end{split}
  \label{eq:delta_v_sphere}
\end{equation}
This formula, also presented in \citet{vanLeeuwen2009}, has been used by e.g.,
\citet{kuhn2019} to `correct' the proper motions for the perspective effect of
the radial velocity.
There are several limitations to such a procedure:
\begin{enumerate}
  \item it requires defining a definite cluster centre.
  \item the mean velocity is estimated from projected velocities,
    which already have the perspective effect baked in.
  \item it is approximate to the first order of $\Delta \alpha$ and $\Delta \delta$.
    While higher-order terms will be smaller, they will always be present.
    They can still be significant, as we are trying to infer the pattern beyond the perspective effect,
    as the astrometric precision becomes better and better.
\end{enumerate}
It is not only conceptually clearer but also more accurate to
forward-model the projection of $\vec{v}_0$ (along with velocity gradient $\mat{T}$)
at each star's position $(\alpha_i,\,\delta_i)$ as we have done in this work.
In doing so, perspective effect is exactly taken into account no matter
how large on the sky the structure is.

Nonetheless, a special case that is worth mentioning is when the velocity is exactly radial:
\begin{equation}
  \vec{v}_\mathrm{sphere} = \left[ 
    \begin{matrix} 0 \\ 0 \\ v_r \end{matrix}
  \right] = \mat{R} \vec{v}_0.
\end{equation}
This may be considered approximately true if, for example,
one decides to subtract mean proper motion of the cluster from the proper motions
of individual stars and look at the residuals, which will mainly be
projections of the mean radial velocity of the cluster (although in reality they will be noisy due to both
observational uncertainties and intrinsic dispersion).
In this special case,  \eqname~(\ref{eq:delta_v_sphere}) simplifies to
\begin{equation}
  \Delta{\vec{v}_\mathrm{sphere}} =
    \left[\begin{matrix}- \Delta\alpha v_{r} \cos\delta \\- \Delta\delta v_{r}\\0\end{matrix}\right]
        + \mathcal{O} (|\Delta \alpha|^2,\, |\Delta \delta|^2,\,|\Delta \alpha \Delta \delta|).
  \label{eq:perspective-radial}
\end{equation}
Because $\cos\delta > 0$ for $-\pi/2 < \delta < \pi/2$,
when $v_r$ is positive (receding cluster), the projected velocities (and proper motions)
becomes more negative with angular distance, i.e., the cluster
is apparently contracting.
The opposite is true when the cluster is approaching leading to a perspective expansion.
This is also intuitively clear as the celestial sphere is
convex with respect to an outward radial vector
and concave to an inward radial vector, as shown in \figname~\ref{fig:perspective-radial}.

In \figname~\ref{fig:perspective-effect}, we illustrate the above equations by
projecting the same velocity vector at three different sky positions.
The velocity is radial at position 1, which results in the perspective contraction pattern
in the first panel of the bottom row, where we show
the projected velocities, $v_\alpha$ and $v_\delta$
at a grid of $(\Delta\alpha,\,\Delta\delta)$.
As discussed above, although the first order linear pattern is well-described by
\eqname~(\ref{eq:perspective-radial}), there are remaining higher-order changes.
As an example, we show how the height of the velocity vector (corresponding to $v_\delta$)
at a slice of $\mathrm{Decl.} = 40$ (indicated with a blue strip)
changes with $\alpha$ in the panel below.
The higher-order terms also depend on $(\alpha,\,\delta)$.
The patterns of the same velocity vary with positions
as can be seen in the latter two panels of the bottom row showing them at position 2 and 3
(at which the same velocity is no longer exactly radial).
Most generally, they are a mix of shear-like (symmetric) and rotation-like (anti-symmetric) patterns.
If the perspective effect is not correctly taken into account, we might
wrongly conclude that a cluster at position 2 is rotating clock-wise and at
position 3 counter-clock-wise when the two clusters have the same velocity
and no real rotation.
In real data, these patterns are further complicated by the depth of the cluster
(differing parallaxes of each star) and
the observational and intrinsic noise (velocity dispersion).

\begin{figure*}
  \centering
  \includegraphics[width=0.95\linewidth]{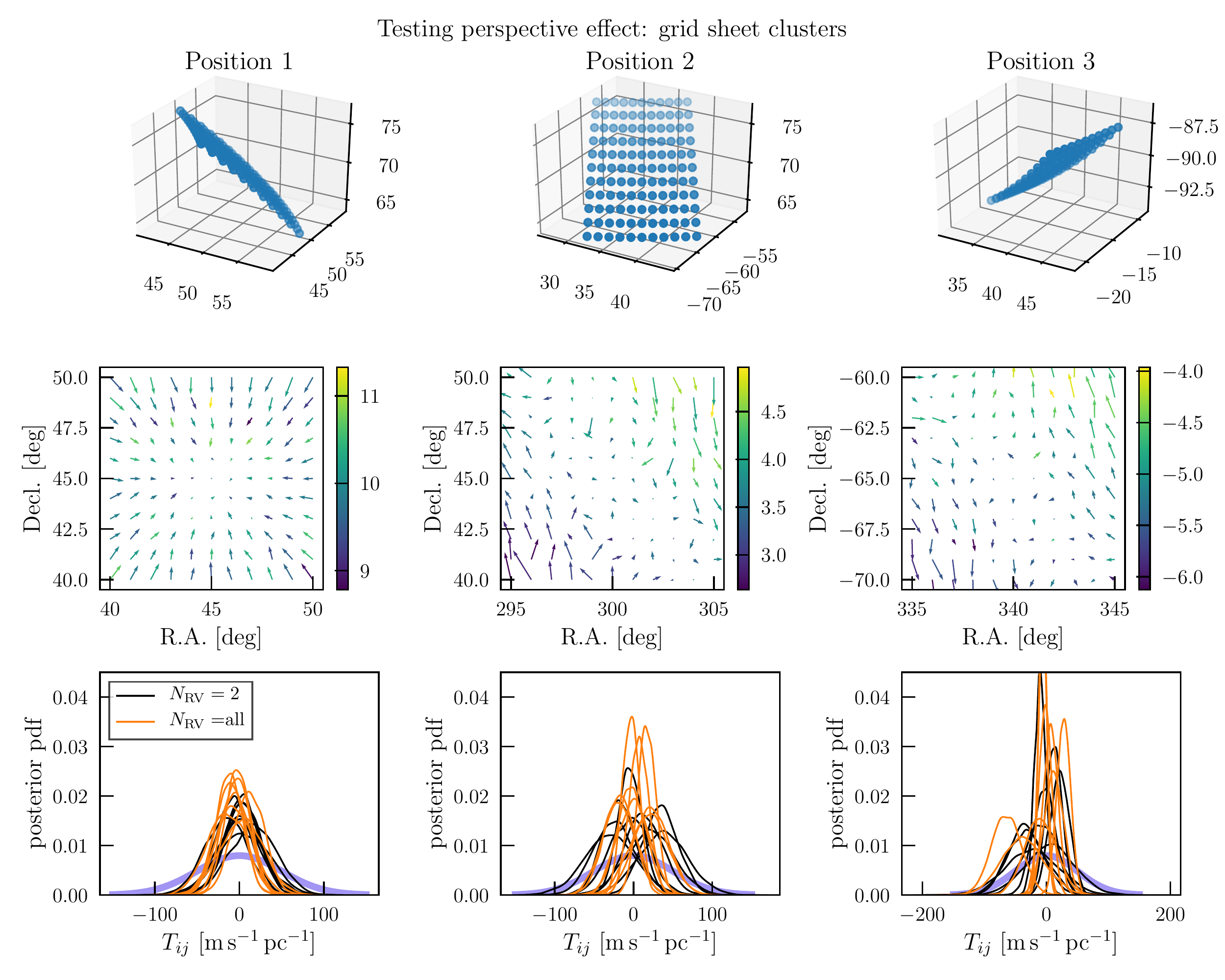} 
  \caption{
    Fitting mock grid sheet clusters at the three positions shown in
    \figname~\ref{fig:perspective-effect}.
    We show the 3D positions of stars in the top row, the residual proper motion
    pattern in the middle row (similar as in
    \figname~\ref{fig:perspective-effect}
    but with observational noise and internal velocity dispersion of 0.1~\kms)
    and the inferred velocity gradient matrix components $T_{ij}=d v_{x,y,z} / d {x,y,z}$.
    In each test case, we assume two scenarios of RV availability:
    when only two randomly chosen stars have RVs and when all stars have RVs.
    The posterior distribution of $T_{ij}$ for these two cases are
    shown in black and orange lines respectively in the bottom panels.
    The proper motion vectors in the middle row are color-coded by radial velocity
    indicated in accompanying right color bars.
    The thick purple line in the bottom panels is the prior distribution of $T_{ij}$.
    The shrinkage from prior to posterior pdf quantifies how informative the data is.
    Note that while the prior of $T_{ij}$ may seem narrow for these test cases
    designed to demonstrate how perspective effect is taken into account,
    they are quite broad for the real Hyades data.
    }
  \label{fig:sheet-clusters}
\end{figure*}

\begin{figure*}
  \centering
  \includegraphics[width=0.95\linewidth]{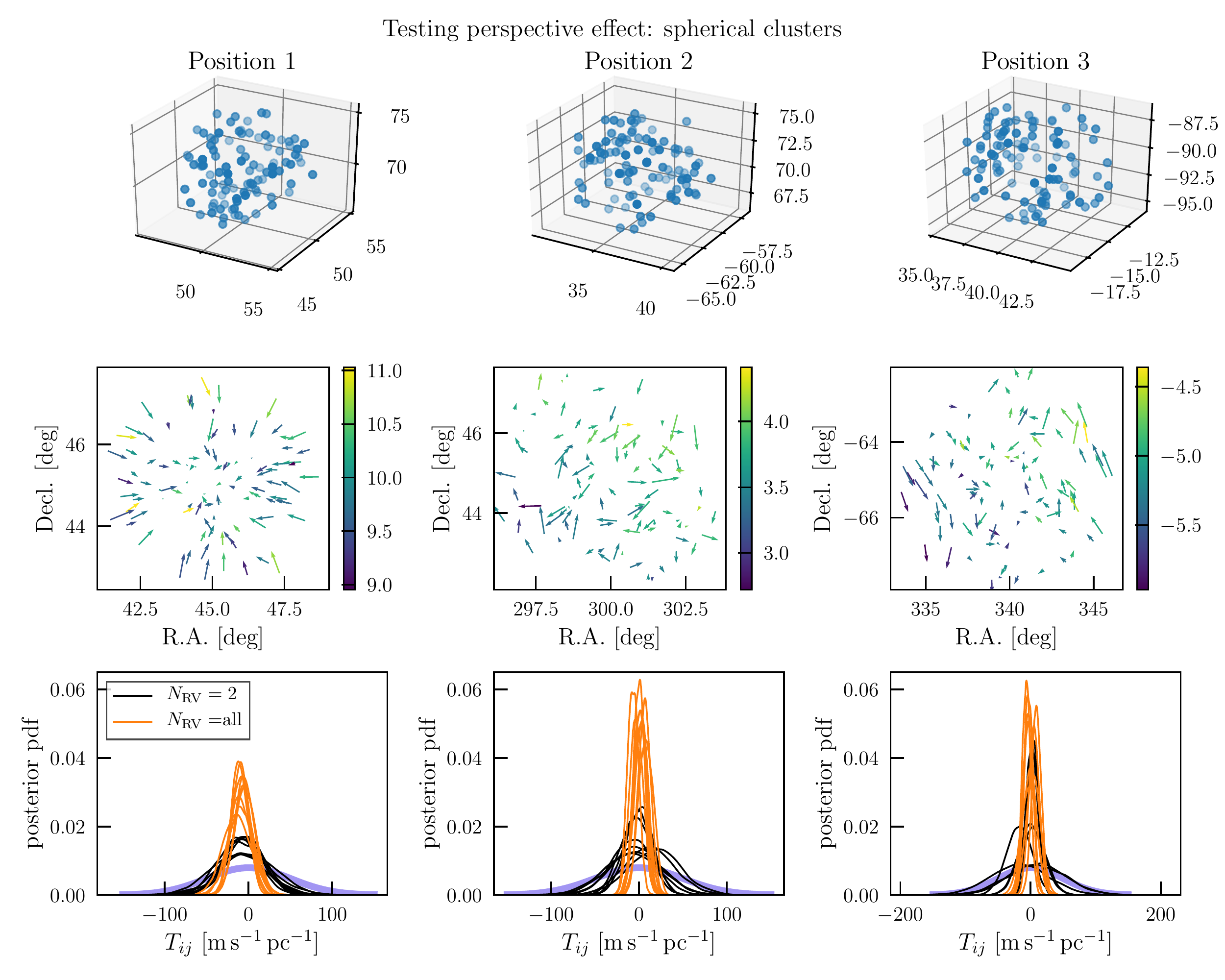} 
  \caption{
    Fitting mock spherical clusters at the three positions shown in
    \figname~\ref{fig:perspective-effect}.
    For details of what each panel shows, see the caption of \figname~\ref{fig:sheet-clusters}.
    The residual proper motion patterns are more messy in these cases as
    they are also affected by each star's distance,
    unlike grid sheet clusters where all stars are at $100$ pc.}
  \label{fig:spherical-clusters}
\end{figure*}

In order to demonstrate the forward-modelling method fully
takes perspective effect into account, we provide two test cases.
First, we fit hypothetical grid sheet clusters sampled at the exact
grids of $(\alpha\,\delta)$ at each position in \figname~\ref{fig:perspective-effect}.
Second, we fit hypothetical spherical clusters centred at each position.
We use a simplified version of our full model, excluding the mixture
component (which accounts for non-member contamination) and assuming
isotropic velocity dispersion to fit the mock data.
We add 5\% Gaussian noise for proper motions and radial velocities,
assume parallax errors small enough to resolve the depth of the clusters,
and give the clusters a small velocity dispersion of 0.1~\kms.
The results are presented in \figname~\ref{fig:sheet-clusters} and
\ref{fig:spherical-clusters}.
In each figure, we show the 3D positions of 100 stars in the mock clusters in the top row,
the observed (sampled with uncertainty) residual proper motion pattern in the middle row
and the inferred posterior pdf of velocity gradient matrix $T_{ij}$ in the bottom row.
For each test case, we assumed two possible scenarios of RV availability:
when RV is available for only 2 stars (black posterior pdf lines)
and when RV is available for all stars (orange posterior pdf lines).
In all cases, we correctly infer $T_{ij}=0$ as we should despite the strong
apparent proper motion gradients (apparent contraction/rotation/shear).
Notice also that while the posterior pdfs of $\mat{T}$ are narrower
when RVs are available for all stars (as they should be with more information),
they are still constrained when RVs are available for only 2 stars.
If a few RV measurements across the cluster can `anchor' the radial velocity dimension, then
subtle changes in proper motions due to the cluster's rotation or shear
can be correctly inferred as they will introduce systematic residual change
on top of the perspective effect.

We conclude this appendix with a final remark that
when the parallax errors are too large such that the depth of the cluster
is unresolved, the inferred velocity gradient can be non-zero even when
its real value is zero.
In this case, the cluster will appear to be radially elongated due to noise
in the parallax measurement, similar to the ``finger of God'' effect for
galaxy clusters.
Because parallaxes give distance information and distances affect proper motions,
this will inject correlations between position and velocity,
which leads to a non-zero inferred $\mat{T}$ even when the real $\mat{T}=0$.
This is a systematic effect, which puts a lower limit on the smallest $\mat{T}$
value that can be inferred from the data.
This is unrelated to taking the covariance between parallaxes and proper
motions into account, which we do in our method described in
\sectionname~\ref{sub:model}, and likely requires a density model
for the cluster along with its kinematics.


\bsp	
\label{lastpage}
\end{document}